\documentclass[structabstract]{aa}

\usepackage{natbib}                     
\usepackage{amsmath}                    
\usepackage{graphicx}                   
\usepackage[varg]{txfonts}              
\usepackage[english]{babel}             
\usepackage{nameref}                    
\usepackage[normalem]{ulem}             
\usepackage{paralist}                   
\usepackage{multirow}

\begin{document}

\title{Broadband transmission spectroscopy of HD\,209458b with ESPRESSO: Evidence for Na, TiO, or both\thanks{Based on Guaranteed Time Observations collected at the European Southern Observatory under ESO programme 1102.C-0744 by the
ESPRESSO Consortium.}}

  \author{N. C. Santos\inst{1,2}
        \and E. Cristo\inst{1,2} 
        \and O. Demangeon\inst{1,2}
        \and M. Oshagh\inst{3,4} 
        \and R. Allart\inst{5} 
        \and S. C. C. Barros\inst{1,2} 
        \and F. Borsa\inst{6} 
        \and V. Bourrier\inst{5} 
        \and N. Casasayas-Barris\inst{3,4} 
        \and D. Ehrenreich\inst{5} 
        \and J. P. Faria\inst{1,2} 
        \and P. Figueira\inst{7,1} 
        \and J. H. C. Martins\inst{1} 
        \and G. Micela\inst{8} 
        \and E. Pall\'e\inst{3,4} 
        \and A. Sozzetti\inst{9} 
        \and H. M. Tabernero\inst{1}
        \and M. R.  Zapatero Osorio\inst{10} 
        \and  F. Pepe\inst{5} 
        \and  S. Cristiani\inst{11} 
        \and R. Rebolo\inst{3,4,12}
        \and V. Adibekyan\inst{1,2} 
         \and C. Allende Prieto\inst{3,4} 
         \and Y. Alibert\inst{13} 
         \and F. Bouchy\inst{5} 
          \and A. Cabral\inst{15,17} 
          \and H. Dekker\inst{18} 
          \and P. Di Marcantonio\inst{11} 
         \and V. D'Odorico\inst{11,14} 
         \and X. Dumusque\inst{5} 
         \and  J. I. Gonz{\'a}lez Hern{\'a}ndez\inst{3,4} 
         \and B. Lavie\inst{3} 
         \and G. Lo Curto\inst{7} 
         \and C. Lovis\inst{5} 
          \and A. Manescau\inst{18} 
         \and C.J.A.P. Martins\inst{1,19} 
          \and D. M{\'e}gevand\inst{5} 
          \and  A. Mehner\inst{7} 
          \and P. Molaro\inst{11,14} 
          \and  N. J. Nunes\inst{15} 
          \and E. Poretti\inst{6,16}  
          \and M. Riva\inst{6} 
          \and S. G. Sousa\inst{1} 
          \and  A. Su{\'a}rez Mascare{\~n}o\inst{3,4} 
          \and  S. Udry\inst{5}
         }

  \institute{
          Instituto de Astrof\'isica e Ci\^encias do Espa\c{c}o, Universidade do Porto, CAUP, Rua das Estrelas, 4150-762 Porto, Portugal
          \and
          Departamento de F\'isica e Astronomia, Faculdade de Ci\^encias, Universidade do Porto, Rua do Campo Alegre, 4169-007 Porto, Portugal
          \and
          Instituto de Astrof\'{i}sica de Canarias (IAC), 38205 La Laguna, Tenerife, Spain
          \and
          Universidad de La Laguna (ULL), Departamento de Astrof\'{i}sica, 38206 La Laguna, Tenerife, Spain
         \and
         Universit\'e de Gen\`eve, Observatoire Astronomique, 51 ch. des Maillettes, 1290 Versoix, Switzerland
         \and
         INAF - Osservatorio Astronomico di Brera, Via Bianchi 46, 23807 Merate, Italy
         \and
         European Southern Observatory, Alonso de C\'ordova 3107, Vitacura, Regi\'on Metropolitana, Chile
        \and
        INAF - Osservatorio Astronomico di Palermo, Piazza del Parlamento 1, 90134 Palermo, Italy
        \and 
        INAF - Osservatorio Astrofisico di Torino, via Osservatorio 20, 10025 Pino Torinese, Italy
        \and
        Centro de Astrobiolog\'\i a (CSIC-INTA), Crta. Ajalvir km 4, E-28850 Torrej\'on de Ardoz, Madrid, Spain
        \and 
        INAF - Osservatorio Astronomico di Trieste, via G. B. Tiepolo 11, I-34143 Trieste, Italy
        \and 
        Consejo Superior de Investigaciones Cient\'{\i}cas, Spain
        \and
        Physics Institute, University of Bern, Sidlerstrasse 5, 3012 Bern, Switzerland
        \and
        Institute for Fundamental Physics of the Universe, Via Beirut 2, I-34151 Grignano, Trieste, Italy
        \and
        Instituto de Astrof\'isica e Ci\^encias do Espa\c{c}o, Faculdade de Ci\^encias da Universidade de Lisboa, Campo Grande, PT1749-016 Lisboa, Portugal
        \and
        Fundaci\'on G. Galilei -- INAF (Telescopio Nazionale Galileo), Rambla J. A. Fern\'andez P\'erez 7, E-38712 Bre\~na Baja, La Palma, Spain
        \and 
        Faculdade de Ci\^encias da Universidade de Lisboa (Departamento de F\'isica), Edif\'icio C8, 1749-016 Lisboa, Portugal
        \and
        European Southern Observatory, Karl-Schwarzschild-Strasse 2, 85748  Garching b. M\"unchen, Germany 
        \and
        Centro de Astrof\'isica da Universidade do Porto, Rua das Estrelas, 4150-762 Porto, Portugal
}

  \date{Received date / Accepted date }
  \abstract
  {The detection and characterization of exoplanet atmospheres is currently one of the main drivers pushing the development of new observing facilities. In this context, high-resolution spectrographs are {proving} their potential and showing that high-resolution spectroscopy will be paramount in this field. }
  {We aim to make use of ESPRESSO high-resolution spectra, which cover two transits of HD\,209458b, to probe the broadband transmission optical spectrum of the planet.}
  {We applied the chromatic Rossiter-McLaughin method to derive the transmission spectrum of HD\,209458b. We compared the results with previous HST observations and {with synthetic spectra}.}
  {We recover a transmission spectrum of HD\,209458b similar to the one obtained with HST data. The models suggest that the observed signal can be explained by only Na, only TiO, or both Na and TiO, {even though none is fully capable of explaining our observed transmission spectrum}. Extra absorbers may be needed to explain the full dataset, though modeling approximations and observational errors can also be responsible for the observed mismatch. }
  {Using the chromatic Rossiter-McLaughlin technique, ESPRESSO is able to provide broadband transmission spectra of exoplanets from the ground, in conjunction with space-based facilities, opening good perspectives for similar studies of other planets.}


\maketitle
  
  \section{Introduction}                                        \label{sec:Introduction}

Following the detection of a planet orbiting the solar-type star 51 Peg b \citep[][]{Mayor-1995}, the field of exoplanet research has not stopped growing. Fostered by the development of new precise Doppler spectrographs \citep[e.g.,][]{Mayor-2014}, space-based transit photometry missions \citep[e.g.,][]{Lissauer-2014}, and exquisite data analysis techniques \citep[see e.g.,][and references therein]{Barros-2020}, the number of known planets is now well over 4000 \citep[e.g., exoplanet.eu --][]{Schneider-2011}. 
Further to the detection of new planets, different methods are now used to routinely derive intrinsic exoplanet properties, including their internal composition and structure \citep[e.g.,][]{Dorn-2019}, as well as their atmospheres \citep[e.g.,][]{Burrows-2014,Sing-2016}. In this latter context, progress has been enormous.

The detection of a giant planet orbiting the bright solar-type star \object{HD\,209458} represents a major cornerstone in exoplanet research. First detected using the radial velocity method, HD\,209458\,b was the first planet to be found to transit its host star \citep[][]{Charbonneau-2000}. This historically confirmed that ``hot jupiters'' were indeed planets, but also allowed a number of complementary studies. These included the first detection of the Rossiter-McLaughlin signature due to a planet \citep[][]{Queloz-2000b}, as well as the first announcement of chemical imprints of the sodium (Na) doublet in the atmosphere of an exoplanet \citep[][]{Charbonneau-2002}. This detection was later backed up (albeit at low resolution, i.e., the Na broad wings were identified) with data obtained with the HST \citep[][]{Sing-2008}. Interestingly, the detection of the core of the Na line in high-resolution transmission spectroscopy was recently questioned by \citet[][]{Casasayas-Barris-2020}, who took into account the previously ignored combination of the center-to-limb variation and the Rossiter-McLaughlin effects to model the data.

The initial atmospheric studies of this planet were followed by a large number of other investigations trying to probe the atmosphere of HD\,209458b using a whole range of techniques and instruments. These include low- and high-resolution transmission spectroscopy, occultation measurements, and phase curves \citep[for these latter ones, see, e.g., the near and mid-infrared (IR) observations of][]{Deming-2005,Knutson-2008,Crossfield-2012}.  Noteworthy results include the detections of Lyman-$\alpha$, oxygen, and carbon in the far-UV \citep[][denoting an evaporating atmosphere]{Vidal-Madjar-2003,Vidal-Madjar-2004}; atmospheric TiO and VO \citep[][]{Desert-2008}; water, methane, and carbon dioxide \citep[][]{Swain-2009,Beaulieu-2010}; calcium \citep[][]{Astudillo-Defru-2013}; magnesium, and ionized iron in the near-UV \citep[][]{Vidal-Madjar-2013,Cubillos-2020};\footnote{Though Mg was not detected by these latter authors.} helium\,I \citep[][]{Alonso-Floriano-2019}; and the signature of Rayleigh scattering \citep[][]{Lecavelier-2008}. Space-based photometry has also shown that HD\,209458\,b is a very low-geometric-albedo planet \citep[][]{Rowe-2008} with a rather high-bond albedo \citep[][]{Schwartz2017}. 

The new ESPRESSO spectrograph \citep[][]{Pepe-2020}, installed at ESO's VLT, offers a new opportunity to study exoplanet atmospheres at high spectral resolution, making use of its unique resolving power and stability, as well as the photon collecting power of the VLT. The first results of ESPRESSO in this context are indeed very promising \citep[][]{Ehrenreich-2020}. The detection of exoplanet atmospheres using high-resolution transmission spectroscopy allows us to study individual absorption lines as well as their precise positions and profiles, thus probing for winds and for the vertical structure of the atmospheres \citep[e.g.,][]{Wyttenbach-2015}. 

Broadband transmission spectroscopy, on the other hand, allows us to probe absorption in exoplanet atmospheres at different wavelength scales, thus providing complementary physical information.
Interestingly, the ultra-precise radial velocities that can be derived with ESPRESSO \citep[see e.g.,][for the case of Proxima\,b]{SuarezMascareno-2020} also open up a new opportunity to study exoplanet atmospheres in this regime.
In this context, in the present paper we investigate the use of ESPRESSO to observe the broadband transmission spectrum of HD\,209458\,b by means of the chromatic Rossiter-McLaughlin effect \citep[][]{Snellen2004,Dreizler-2009,DiGloria-2015}. In Sect.\,\ref{sec:data}, we present our dataset, and in Sect.\,\ref{sec:model} we present the model used. We then present and discuss the results in Sects.\,\ref{sec:transmission} and \ref{sec:discussion}. We conclude in Sect.\,\ref{sec:conclusions}.

  \section{Data}                                \label{sec:data}

Two transits of HD\,209458\,b were observed with the ESPRESSO spectrograph \citep[][]{Pepe-2010,Pepe-2020} at the VLT on the nights of 2019/07/19 and 2019/09/10 (during guaranteed time observation run 1102.C-0744). A total of 174 (89 + 85) spectra were obtained and reduced using version 2.0.0 of the ESPRESSO data reduction system (DRS). Exposure times were set to 175 seconds. All observations were obtained with the UT3  telescope in the HR21 mode. 

\begin{figure*}[t!]
\begin{center}
\begin{tabular}{c}
\includegraphics[angle=0,width=0.45\linewidth]{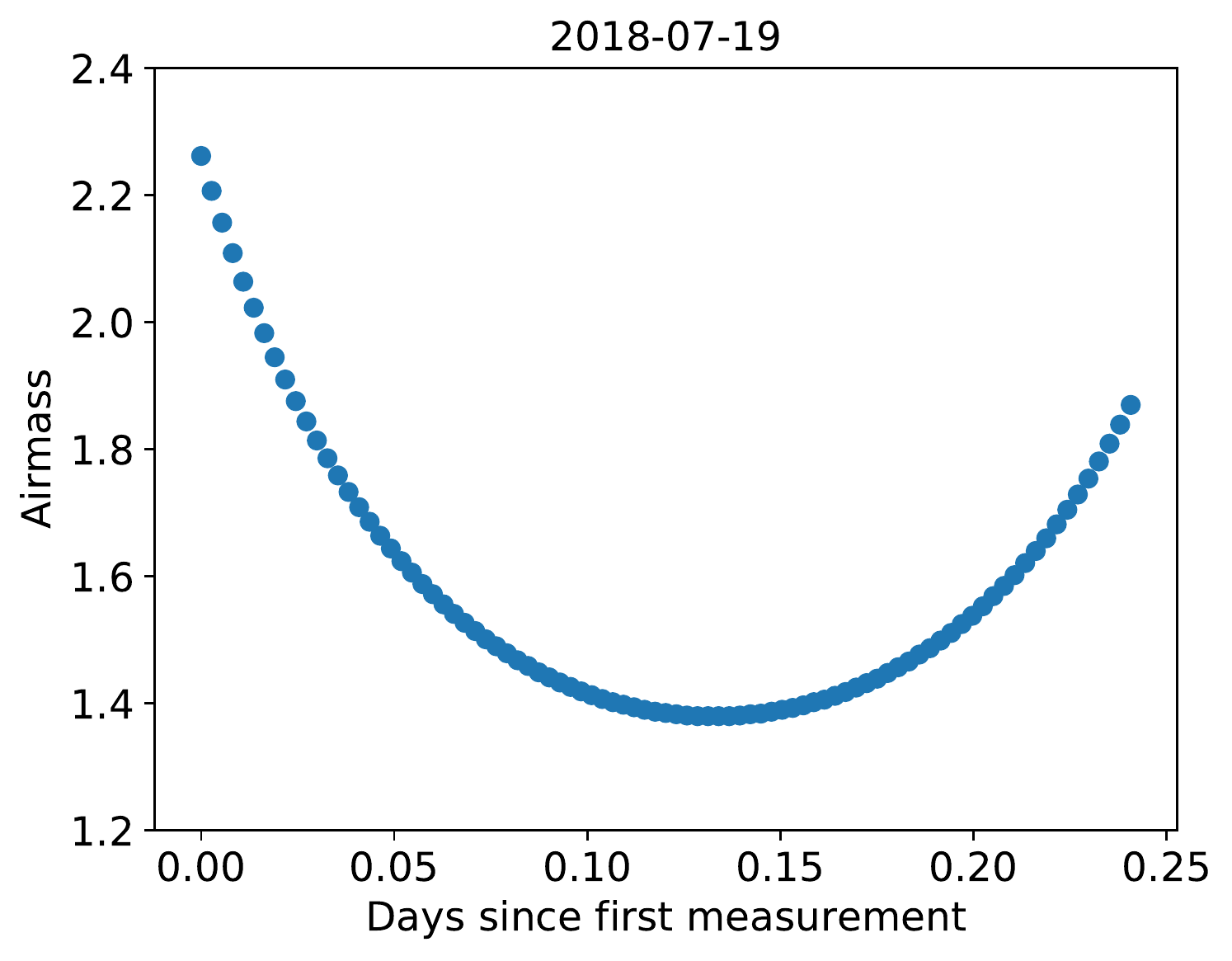}
\includegraphics[angle=0,width=0.45\linewidth]{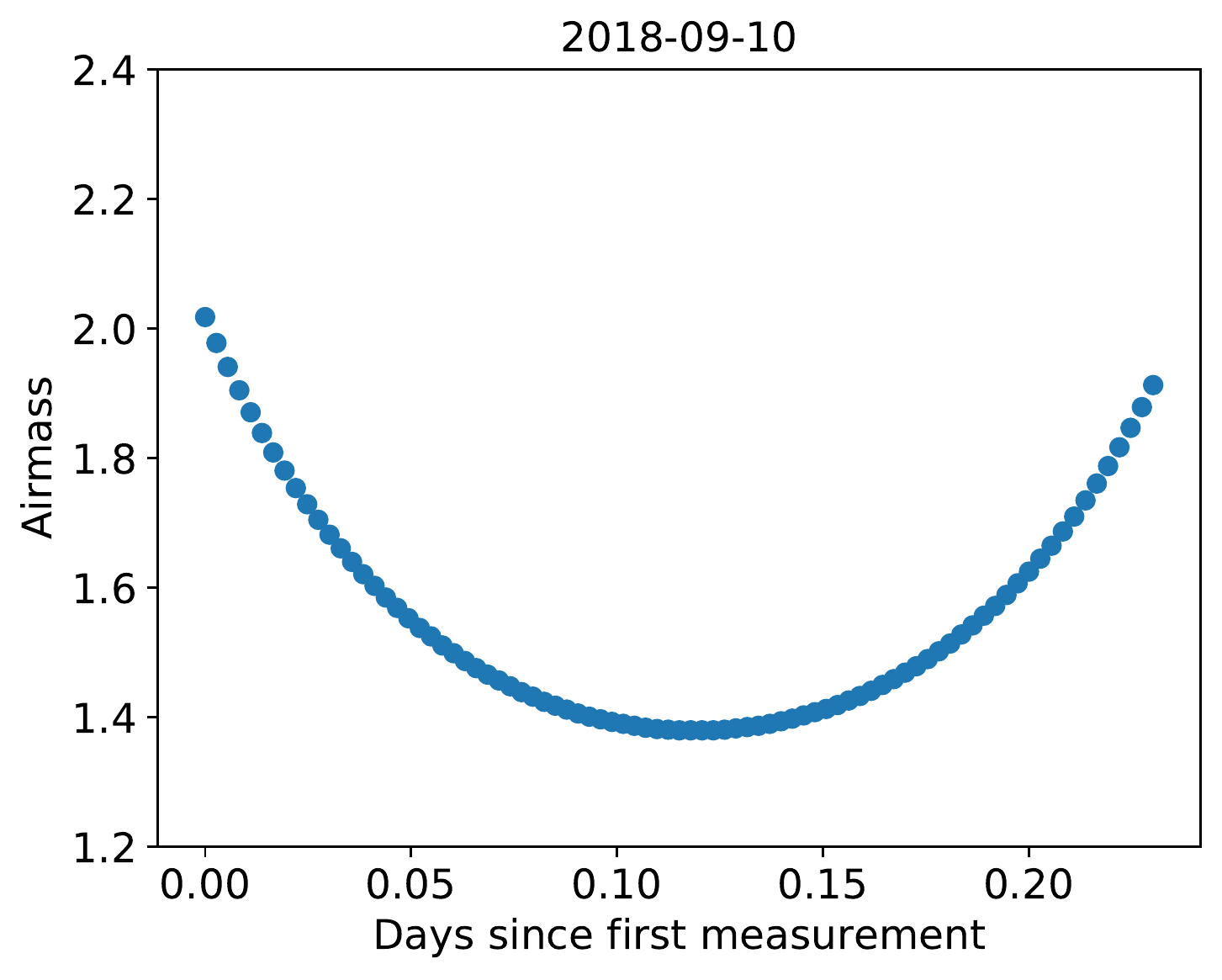}\\[-0.0truecm]
\includegraphics[angle=0,width=0.45\linewidth]{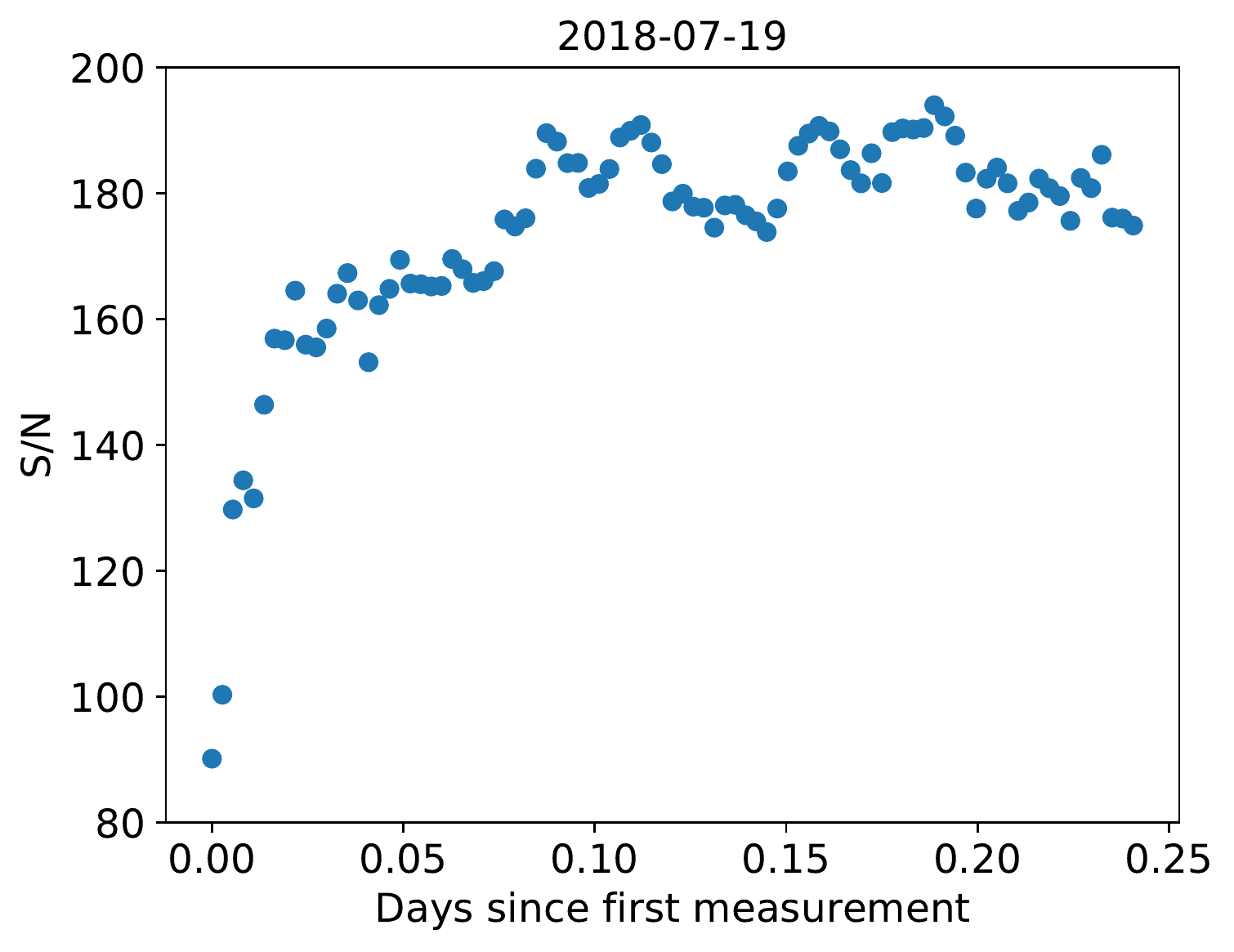}
\includegraphics[angle=0,width=0.45\linewidth]{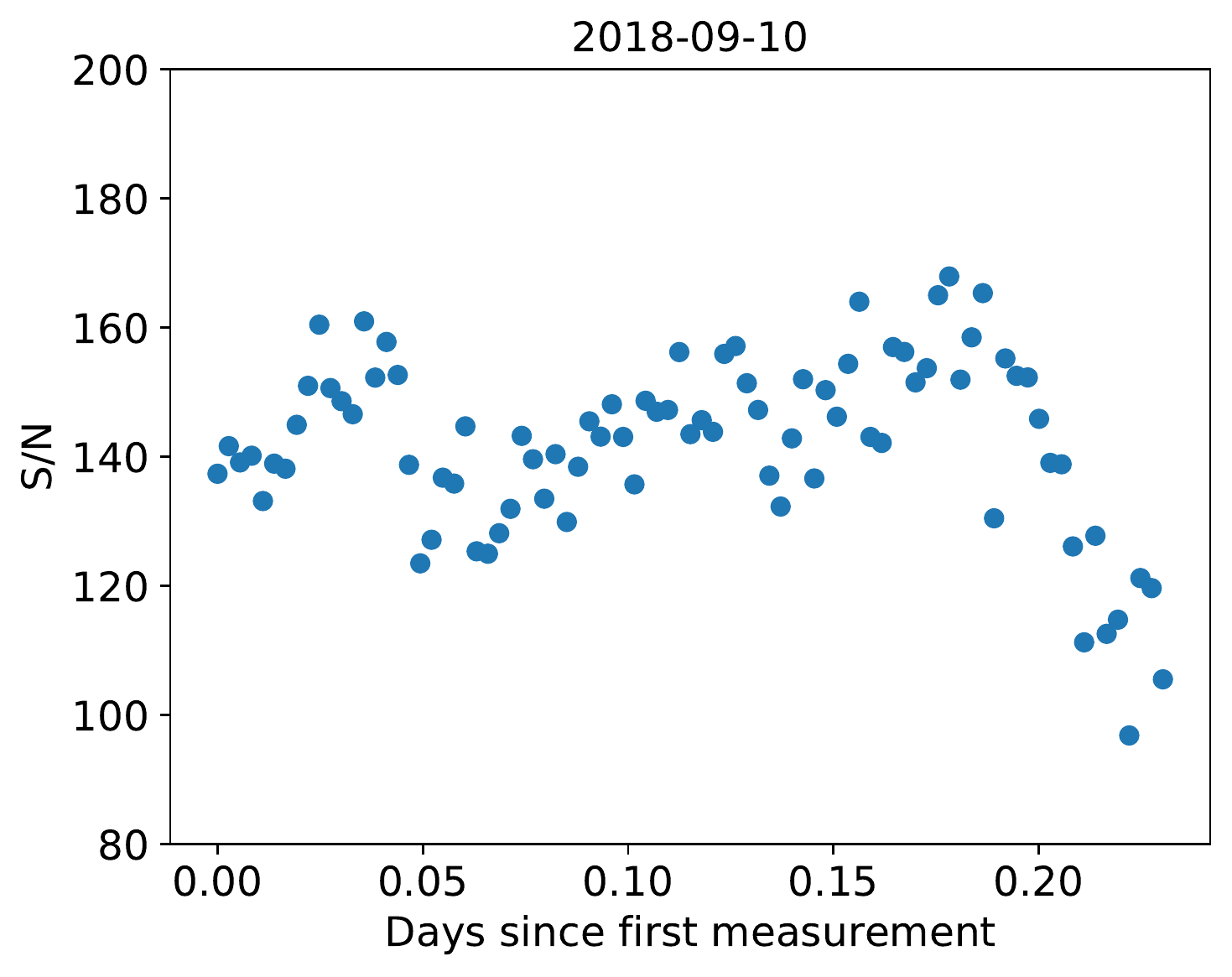}\\[-0.0truecm]
\end{tabular}
\end{center}
\vspace{-0.5cm}
\caption{Times series of air mass and S/N ratio near 580\,nm for the two transits. Data for the first and second transits are shown in the left and right plots, respectively.}
\label{fig:obs}
\end{figure*}

Each obtained spectrum covers the full wavelength domain from 380\,nm to 788\,nm at a resolution of $\lambda/\Delta\lambda\sim140\,000$. The spectrum is divided into 85 echelle interference orders, but due to the optical design of ESPRESSO, each order is repeated twice (in practice we see a total of 170 ``slices''). Typical signal-to-noise (S/N) ratios per pixel near 580\,nm (as measured in the spectral slice 116) are 180 and 140, respectively, for the first and second transits. This leads to typical radial velocity (RV) uncertainties of the order of 50 cm/s. {In Fig.\,\ref{fig:obs}, we present the variation of the S/N (for spectral slice 116) and air mass during the observations, and in Fig.\,\ref{fig:WL} (top panels) we show our RV time series for the two nights.}

The ESPRESSO DRS provides the cross-correlation function (CCF) for each spectral slice. In our case, we used the DRS-standard correlation mask built from a spectrum of an F9V star, the same spectral type as HD\,209458 (F9V). Given the brightness of our target, we used the "non-sky subtracted" spectrum. The RVs derived from the sky-subtracted spectrum are in fact similar (within the errors), and we verified that the use of that version does not change any of the results of this paper. Slices where telluric lines heavily affect the spectrum are masked, and their CCF is not computed.\footnote{This includes slices 118/119, 136/137, 148/149 through 156/157, and 164/165. We note that, for convenience, we define slices 0/1 as the bluest, and 168/169 the reddest.}

\section{The model}                             \label{sec:model}

This work is based on the use of the Rossiter-McLaughlin (RM) effect \citep[][]{Rossiter1924,McLaughlin1924} to study how the planet radius varies as a function of wavelength, using a similar approach to the one used by \citet[][]{DiGloria-2015} to study the transmission spectrum of \object{HD\,189733b}. This implementation of the RM technique is also known as the chromatic Rossiter-McLaughin. The rationale is based on the fact that the RM signal (namely its amplitude) depends strongly on the planet-to-star radius ratio. As such, if the planet radius varies as a function of wavelength, we should be able to detect its variation by modeling the RM curve for different spectral bands \citep[][]{Snellen2004,Dreizler-2009}.

We used the CARM code to fit the chromatic RM effect on HD209458. A detailed description of the code will be presented in a separate paper (Cristo et al., in prep.). Briefly, the RV is modeled as the sum of a sinusoidal function (i.e., a circular ``Keplerian function'' to take into account the RV variation along the planet orbit)\footnote{The circular orbit approximation is valid for HD\,209458, as it is for the large majority of known hot jupiters. We nevertheless do not expect a possible small eccentricity value to cause errors in our analysis, since it is not a color-dependent parameter.} and an RM curve. In our specific case, the choice of a ``zero-eccentricity'' orbit is perfectly justified by the known orbital properties of the planet \citep[e.g.,][]{Southworth2010}. The code allows us to choose between two approaches to fit an RM model to a radial velocity time series: the one implemented in {\sc Arome} \citep[][]{Boue-2013}, and the one implemented in {\sc PyAstronomy} \citep[][]{pya}\footnote{https://github.com/sczesla/PyAstronomy} and based on the model of \citet[][]{Ohta-2005}. One of the relevant differences between them is the treatment of the limb darkening: while {\sc Arome} allows for both quadratic and nonlinear limb-darkening laws, in the model implemented in {\sc PyAstronomy} only a linear limb darkening law is allowed. When fixing the limb darkening values we use the {\sc ldtk} package \citep[][]{Parviainen-2015}, that computes the limb darkening coefficients (quadratic or linear) based on the provided effective temperature, surface gravity, and metallicity of the star (and respective uncertainties), and the wavelength interval of interest.

The ``Keplerian'' part of the model is defined with the following parameters: the systemic velocity ($V_{sys}$) of the star (the ``zero point'' of the radial velocity of the Keplerian orbit), the semi-amplitude ($K$), mid-transit time ($T_0$), and the orbital period ($P$). For the RM model, the input parameters depend on the formalism used. In {\sc Arome}, these include the (quadratic) limb-darkening parameters, the orbital inclination $i_p$, the projected spin-orbit angle ($\lambda$), the planet-to-star radius ratio ($R_p/R_\star$), the semi-major axis of the orbit ($a$), the stellar projected rotational velocity ($v\,\sin{i}$), the stellar macro-turbulence ($\zeta_t$), as well as the spectroscopic parameters $\sigma_0$ and $\beta_0$, which denote the width of the observed out-of-transit CCF (in km\,s$^{-1}$) and of the CCF for a nonrotating star with the same physical properties, respectively. 
When using the model implemented in {\sc PyAstronomy,} the latter four parameters ($v\,\sin{i}$, $\zeta_t$, $\sigma_0$, $\beta_0$) are replaced 
by the angular velocity of the star ($\Omega$; or alternatively, and as used in our case, the rotational velocity, $V_{rot}$; these are connected through a fixed stellar radius) and the stellar inclination ($i$). In the present version of the code, except for the orbital period $P$, all the remaining parameters can be adjusted or fixed during the fit, together with an additional optional ``jitter'' term ($\sigma_w$).

The CCF files computed by the ESPRESSO DRS are used as input in CARM, together with the slice intervals (corresponding to wavelength intervals) for which we want to apply our model. Based on those, for each CCF file, the code sums the CCFs in that specific slice interval. A Gaussian fit to the resulting CCF is then used to retrieve the radial velocity. The uncertainty is also computed using the method described in \citet[][]{Bouchy-2001}. This approach is also used by the DRS to compute the stellar radial velocity from the fully combined, white-light (summed over all slices) CCF, as well as its uncertainty. At the end of this step, we have one RM time series for each wavelength region (slice interval) of interest. These time series are then fit using one of the models mentioned above using the Markov chain Monte Carlo (MCMC) package {\tt emcee} \citep[][]{Foreman-Mackey-2013}.

We note that the modeling of the (chromatic) Rossiter-McLaughlin effect does not require the data to be flux calibrated. This is a huge advantage with respect to photometric techniques.

\section{The transmission spectrum of HD\,209458b}
\label{sec:transmission}

\subsection{The fit to the white light RM}

Modeling the RM effect implies varying several parameters. Some of them, however, are expected to be color independent. As such, with CARM, we started by fitting the (higher S/N) white-light RVs, that is, the RVs obtained by summing all the CCFs in the full spectrum. These RVs are also used to measure the stellar RV in a given epoch (this is the value that is provided directly by the ESPRESSO DRS). This allows us to fix some of the fit parameters when later adjusting each individual RM curve for a given wavelength regime.

\begin{table*}[t]
\caption{Reference values for the model parameters. }
\begin{center}
\begin{tabular}{lll}
\hline
Parameter      & Value & Source\\
\hline
$R [R_{\odot}]$ & $1.162^{+0.012}_{-0.008}$ &  \citet{Southworth2010}\\
$T_{eff} [K]$ & $6118\pm25$& \citet{Sousa2008}\\
$log(g)$ & $4.36\pm0.04$& \citet{Sousa2008}\\
$[Fe/H]\,[dex]$ & $0.03\pm0.02$& \citet{Sousa2008}\\
$T_0\,[BJD]$ & $2455216.405640\pm0.000094$& \citet{Zellem2014}\\
$P [days]$ & $3.52474859\pm0.00000038$ & \citet{Zellem2014}\\
$a [A.U]$ & $0.04747^{+0.00046}_{-0.00031}$ & \citet{Southworth2010}\\
$\lambda [^\circ]$ & $-5\pm7$ & \citet{Albrecht2012}\\
$v\,sin(i_\star) [Km/s]$ & $4.2\pm0.1$& Casasayas-Barris et al., in prep.\\
$V_{sys} [Km/s]$ & $-14.78$ &  See text\\
$R_p [R_{Jup}]$ & $1.380^{+0.015}_{-0.009}$ & \citet{Southworth2010}\\
$K [m/s]$ & $ 84.67\pm0.70$ &\citet{Torres2008}\\
$i [^\circ]$ & $ 86.71\pm0.05$ &\citet{Torres2008}\\
$\sigma_0 [Km/s]$ & $ 4.06$ & This work\\
$\beta_0 [Km/s]$ & $3.2$ & This work\\
$\zeta_t [Km/s]$ & $4.39\pm0.01$ &From the calibration of \citet{Doyle2014}\\
\hline
\hline
\end{tabular}
\end{center}
\label{tab:paramWL}
\end{table*}%

As mentioned above, the full model is comprised of the sum of a Keplerian function and an RM curve. For the latter, we decided to use the {\sc Arome} code as the baseline, since it allows for the correct modeling of the RVs as derived using the CCF method \citep[see][]{Boue-2013}, as well as the use of a quadratic limb-darkening law. The baseline parameters used in this fit, together with their sources, are listed in Table\,\ref{tab:paramWL}. The priors used are listed in Table\,\ref{tab:priorsWL}. All parameters not listed were fixed to the values in Table\,\ref{tab:paramWL}.

Some parameters deserve a special note. The value of $V_{sys}$ depends on the spectrograph and even on the wavelength used. As such, we guessed its value from the values of the observed RV data. The prior we used (see below) is broad enough to be mostly uninformative and allow any uncertainty. The value of $\sigma_0$ we adopted is the average width of the observed "out-of-transit" CCFs (4.03 km/s). The central value adopted for $v\,\sin{i}$ (4.2\,km/s) was taken from a detailed analysis of the ESPRESSO data presented in Casasayas-Barris et al. (in prep.).\footnote{Based on the reloaded Rossiter-McLaughlin method \citep[e.g.,][]{Cegla-2016}.} The value is actually half-way between the values derived by 
\citet[][$3.75\pm1.25$\,km/s]{Queloz-2000b}, \citet[][$4.4\pm0.2$\,km/s]{Albrecht2012}, and \citet[][$4.70\pm0.16$\,km/s]{Winn-2005}. The value of $\beta_0$, which represents the width of the CCF for a nonrotating star, is difficult to derive. As such, we decided to
leave it as a free parameter to be fit (see prior discussion below). The macroturbulence parameter ($\zeta_t$) was derived from the calibration of \citet{Doyle2014}, based on the temperature and surface gravity as listed in Table\,\ref{tab:paramWL}. We note that $\beta_0$, $\sigma_0$, and $\zeta_t$ are degenerate in the model. Changing their values (individually) would change the measured planet radius. They are, however, supposed to be mostly wavelength independent.\footnote{We measured, for example, a maximum variation of $\sigma_0$ of only 0.03 km/s from slice to slice.} Since in this work we are ultimately interested in variations of radius as functions of wavelength, the zero point of the radius is not relevant. As such, the exact values for these parameters are not critical. Finally, the value of $a$ used as input in {\sc Arome} is actually given in stellar radius units. We use the values in Table\,\ref{tab:paramWL} to compute it.

The limb-darkening parameters were set using the {\sc ldtk} package, and assuming the following stellar parameters listed in the Sweet-Cat database \citep[][]{Santos-2013,Sousa-2018}: $T_{eff}=6118\,K\pm25$, $\log{g}=4.36\pm0.04$, [Fe/H]=0.03$\pm0.02$ \citep[their original source is][]{Sousa-2008}. The derived values for the linear and quadratic parameters are 0.56 and 0.12, respectively.

For the priors (Table\,\ref{tab:priorsWL}), the values used are based on Table\,\ref{tab:paramWL}. For $K$ and $R_p,$ we assumed Gaussian priors with widths three times larger than the listed errors. This takes into account errors in $K$ caused by activity-related noise (not taken into account when fitting the Keplerian) and the fact that the planet radius comes from observations in a different band.
For  $i$, $\lambda$, and $a$, we used the values as provided in the literature (Gaussian distributions around the provided value); these values are not supposed to depend on our observational method. 
For $V_{sys,}$ we assumed a uniform prior with a broad width of 0.5\,km/s. For the mid-transit time, we propagated the error in $T_0$ and $P$ using the prescription of \citet[][]{Kane-2009}. The prior on $T_0$ is given by an error margin in phase, $\Delta \phi_0$. 

The value of $\beta_0$ is more difficult to estimate, since it is not possible to observe it directly. We thus estimated the starting value based on a comparison between the observed width of the CCF in HARPS (3.61\,km/s) and the ESPRESSO (4.03\,km/s) data for HD\,209458. Using a similar approach as the one presented in \citet[][]{Santos-2003b}, $\beta_0$  was then estimated as 2.6\,km/s for HARPS. Accounting (quadratically) for the excess CCF width of the ESPRESSO vs. HARPS,\footnote{Though ESPRESSO has higher spectral resolution, the ESPRESSO CCFs are wider than HARPS, due to a change in the way lines are weighted when computing the CCF.} this allows us to derive a value of $\beta_0$ for ESPRESSO of  3.2\,km/s. 
In the model, we adopted a very large, uniform prior around this value ($\pm$ its absolute value).

\begin{figure*}[t!]
\begin{center}
\begin{tabular}{cc}
\includegraphics[angle=0,width=0.45\linewidth]{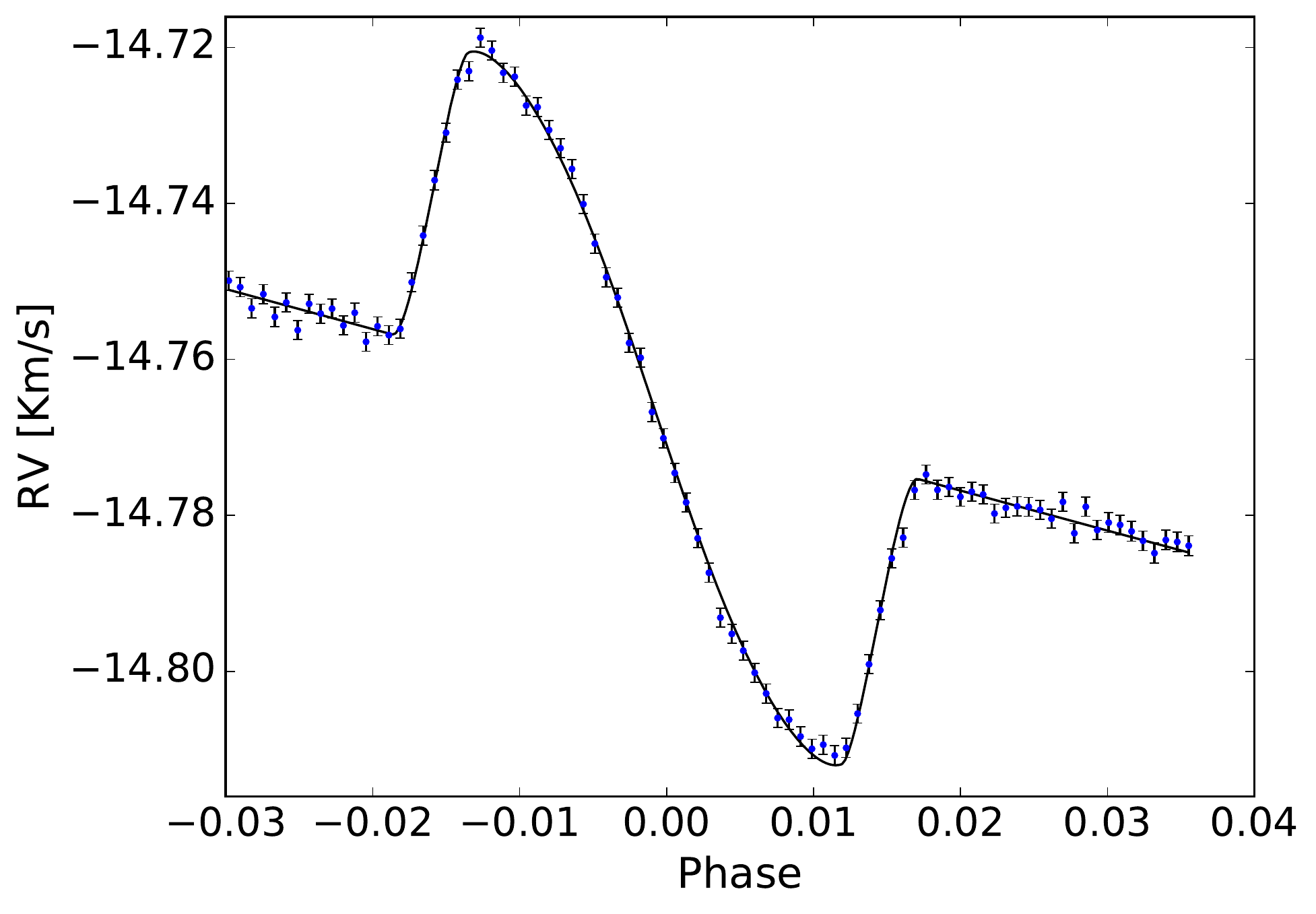} & \includegraphics[angle=0,width=0.45\linewidth]{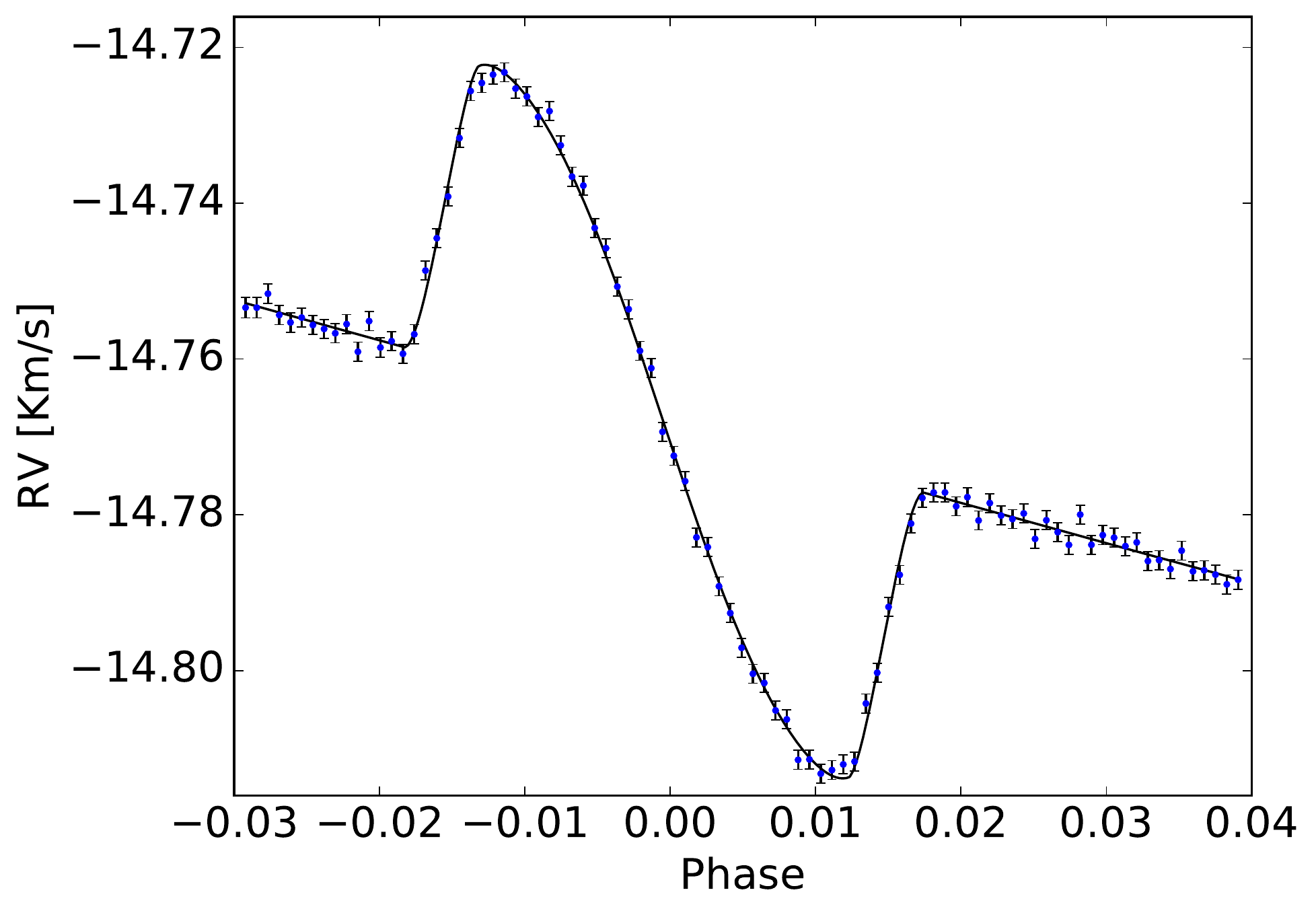} \\
\includegraphics[angle=0,width=0.45\linewidth]{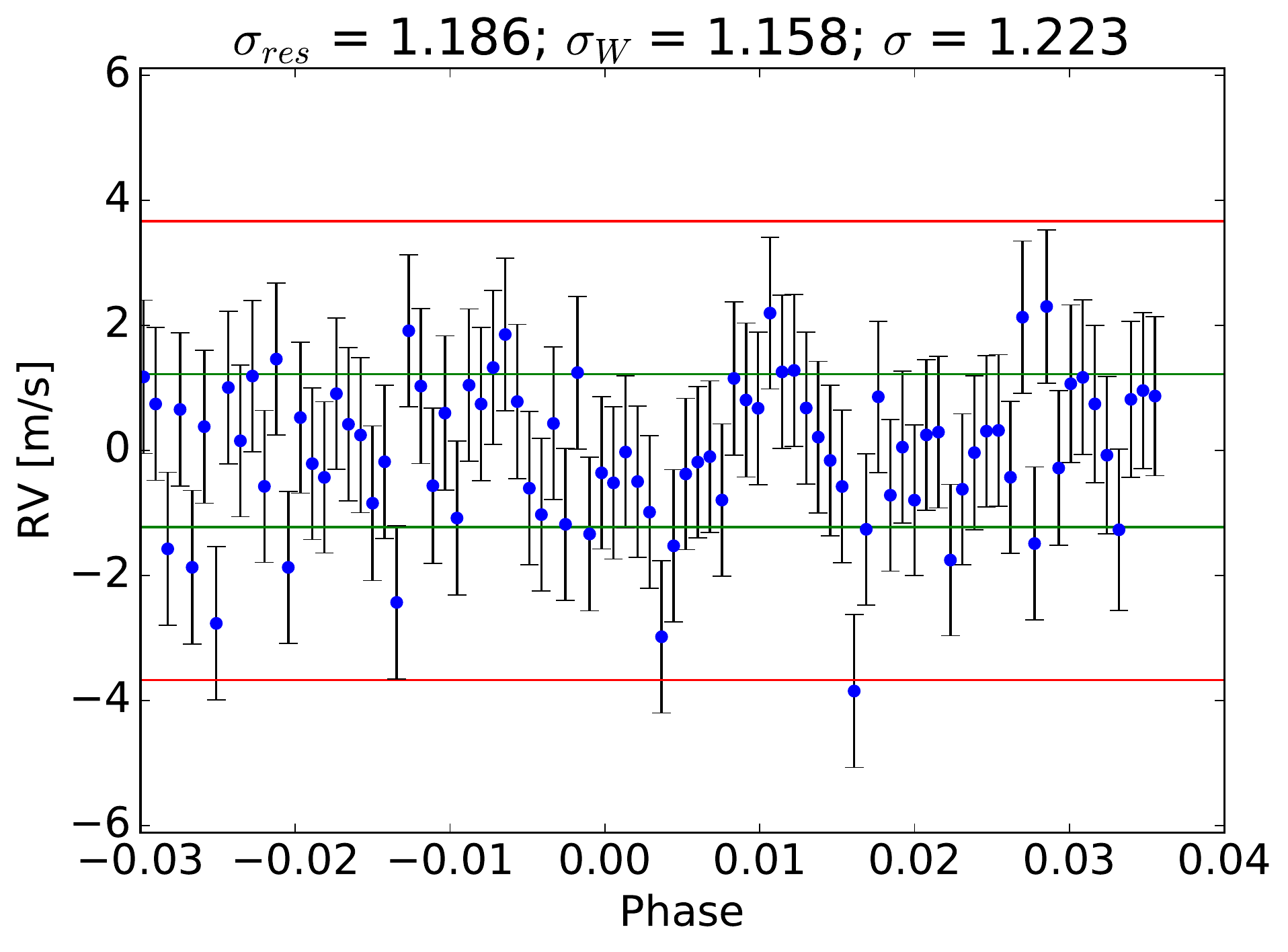} & \includegraphics[angle=0,width=0.45\linewidth]{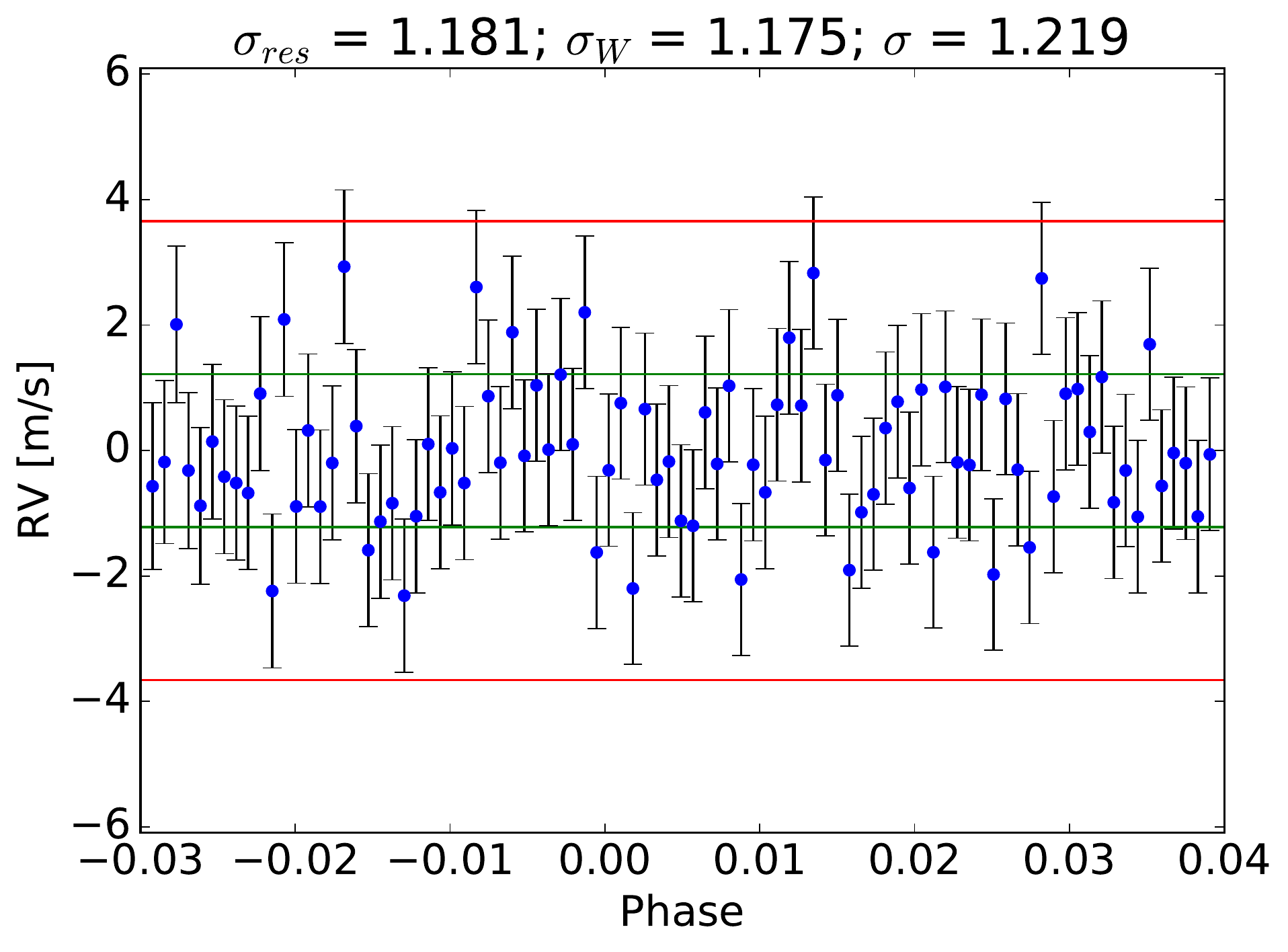} \\
\end{tabular}
\end{center}
\vspace{-0.5cm}
\caption{Rossiter-McLaughlin curves and best fit models (upper panel), as well as residuals (lower panels) for the two separate nights (left and right for the first and second transits, respectively). Plots are shown as functions of orbital phase. The sigma values presented on top of each of the lower panels denote the dispersion of the residuals, the fit jitter, and the quadratic sum of the jitter with the average photon noise error, respectively. }\label{fig:WL}
\end{figure*}

Finally, for the additional ``jitter'' term, we assumed values between 0.0 and 2.5\,m/s. The latter value is five times the observed average photon noise error. This upper limit should take into account any noise coming from stellar oscillations and granulation \citep[e.g.,][]{Dumusque-2010}, on top of additional instrumental noise that for ESPRESSO are expected to reach $\sim$10 cm/s \citep[][]{Pepe-2010}.

We note that although the two transits were fit simultaneously, $V_{sys}$, $\Delta \phi_0$, $K$, and $\sigma_W$ were fit independently in the two cases. This allows us to take into account potential variations of these parameters due to, for example, stellar activity effects (it is, e.g., through $K$ that we can model longer term trends caused by stellar noise). {The resulting values have proven to be similar between the two epochs, even if (as expected) some differences exist. In particular, the values of K and $\log{\sigma_W}$ are indistinguishable in both cases ($0.083\pm0.001$\,km/s and $-6.76\pm0.09$ vs. $-6.75\pm0.09$, respectively), while $V_{sys}$ shows some variation ($-14.7661\pm0.0002$ vs. $-14.7678\pm0.0002$ km/s). The phase shift $\Delta \phi_0$ is also slightly different: $0.00078\pm0.00004$ vs. $0.00042\pm0.00003$}. 

\begin{table*}[t]
\caption{Priors used in the white-light fit using {\sc Arome} as well as the posterior values found. When two values for the posterior are presented, they correspond to the first and second transits, respectively.  {$\mathcal{U}$ and  $\mathcal{G}$ denote uniform and Gaussian prior distributions, respectively.}}
\begin{center}
\begin{tabular}{lll}
\hline
Parameter      & Prior  & Posterior \\
\hline
$V_{sys} [Km/s]$ & $\mathcal{U}(-15.28,-14.28)$ & $-14.7660\pm0.0002/-14.7678\pm0.0002$\\ 
$R_p /R_\star $ & $\mathcal{G}(0.1193,0.0039)$ & $0.120\pm0.001$\\ 
$K [m/s]$ & $\mathcal{G}(84.67,2.10)$ & $83\pm1/83\pm1$ \\ 
$a /R$ &  $\mathcal{G}(8.78,0.26)$ & $8.90\pm0.03$\\ 
$i [^\circ]$ & $\mathcal{G}(86.71,0.05)$ & $86.66\pm0.05$\\ 
$\lambda [^\circ]$ &  $\mathcal{G}(-5,7)$ & $0.6\pm0.4$\\ 
$\beta_0 [Km/s]$ &  $\mathcal{U}(0,6.4)$ & $2.4\pm0.3$\\
$v\,sin(i_\star) [Km/s]$ &  $\mathcal{G}(4.2,0.2)$ & $4.8\pm0.2$ \\
$\Delta \phi_0$ & $\mathcal{G}(0.00,0.02)$ & $0.00078\pm0.00004/0.00042\pm0.00004$\\ 
$\sigma_W$ [m/s] &  $\mathcal{U}(0, 2.5)$ & $1.2\pm0.1/1.2\pm0.1$\\
\hline
\hline
\end{tabular}
\label{tab:priorsWL}
\end{center}
\end{table*}

The result of the RM fit to the two transits is presented in Fig.\,\ref{fig:WL}. The MCMC was run using a total of 60 walkers (or chains) taken randomly from the priors, allowing them to run for 2000 steps (burn-in) before continuing into the production chains (2000 steps). 
{The fit of the {\sc Arome} model} is computationally intensive.\footnote{Using a server with 24 cores@2.3GHz  and 128 GB RAM: four to five hours for the analysis of the white-light RM curve and a few days for the chromatic analysis performed in Sect.\,\ref{sec:chromatic}.} 
For some computations, the length of the burn-in phase was insufficient for all the walkers to converge toward the maximum. To avoid non-converged walkers biasing the posterior distributions, we discarded walkers whose posterior value was abnormally lower than that of the rest of the walkers.\footnote{For this purpose, we used the following criteria: $\ln{post_{walker_i}}<median(\ln{post_{walker_{i...N}}}) - 3\times\,MAD(\ln{post_{walker_{i...N}}}$), where $\ln{post_{walker_i}}$ is the logarithm of the posterior of walker $i$, and $median(\ln{post_{walker_{i...N}}}) $ and $MAD(\ln{post_{walker_{i...N}}})$ are the median and median absolute deviation of all walkers, respectively.} Using this approach, we rejected 15 of the 60 walkers from our analysis of the white-light RM curve.
The posterior distributions, built from the remaining 90\,000 iterations, are presented in Fig.\,\ref{fig:cornerWL}. This plot shows that all parameters converged to distributions that are narrower than the priors.

The median values and their 68\% confidence intervals are presented in Table\,\ref{tab:priorsWL}. A few of the parameters deserve some comments. First, $v\,\sin{i}$ converged to a high value ($4.8\pm0.2$\,km/s), three-sigma away from the best literature estimate ($4.2\pm0.1$\,km/s). As shown in \citet[][]{Brown2017}, there is usually a disagreement between the $v\,\sin{i}$ values derived from the RM fit when compared to those derived from detailed spectroscopic analysis. On its side, $\beta_0$ converged toward a lower-than-expected value. As can be seen in Fig.\,\ref{fig:cornerWL}, and as expected, $\beta_0$ and $v\,\sin{i}$ are highly correlated. 

The derived value of $\lambda=0.6\pm0.4^\circ$ (the spin-orbit angle) is closer to the $1.58\pm0.08^\circ$ derived by Casasayas-Barris et al. (in prep.), based on the reloaded Rossiter-McLaughlin method \citep[][]{Cegla-2016}, than to other literature estimates: $-4.4\pm1.4^\circ$ \citep[][]{Winn-2005} and $-5\pm7^\circ$  \citep[][]{Albrecht2012}.

Given that $\zeta_t$ is also expected to be strongly correlated with $\beta_0$ and $v\,\sin{i}$, we decided to run the MCMC code again, but to also leave $\zeta_t$ as a free parameter (using a very broad, uniform prior between 0 and 8.8, which is twice the expected value). As a result, we obtained values of $\beta_0$, $v\,\sin{i}$, and $\zeta_t$ of 
$3.5\pm0.3$\,km/s, $4.6\pm0.2$\,km/s, and $1. 3^{+1.2}_{-0. 9}$\,km/s, respectively. The values for all the remaining variables were, within the error bars, similar to the ones presented in Table\,\ref{tab:priorsWL}. We thus conclude that the exact input values for $\beta_0$, $v\,\sin{i}$, and $\zeta_t$ are strongly correlated and will not significantly affect the results. We also confirmed that this approach did not change any of the results presented in the next section (i.e., that using a free $\zeta_t$ did not alter the obtained transmission spectrum).

A look at the residuals of the fit to the white-light data (Fig.\,\ref{fig:WL}, lower panels) shows some structure, with a peak-to-peak amplitude of $\sim$2\,m/s. In principle, this may be related to uncertainties in the limb-darkening law or in the RM model, as well as to stellar noise produced by, for example, activity crossing events and stellar pulsation. In \citet[][]{DiGloria-2015}, the authors corrected this structure using the white-light curve as a proxy for the signal in the different wavelengths. By doing so, they assumed that their residual structure was color independent. Without fully understanding the cause for the observed trends, we preferred, however, not to apply any correction.

\subsection{The Chromatic RM}
\label{sec:chromatic}

The next step in our analysis is the derivation of the planet-to-star radius ratio as a function of wavelength. This was done by using the same model as explained in the previous section, but by fitting it to each wavelength interval.
In this process, we fixed all parameters that were returned from the white-light RM fit (Table\,\ref{tab:priorsWL}), except the systemic velocity of the star,\footnote{Due to line-formation depths and convective motions, different lines and spectral regions are expected to have different measured RV values.} the noise ``jitter'' term, and the planet-to-star radius ratio. The priors used for these three parameters are presented in Table\,\ref{tab:priorsChrom}. For $V_{sys,}$ we assumed a uniform prior with a width of 0.5\,km/s around the value from the white-light fit. For $R_p/R_\star$, we used a Gaussian prior centered in the value derived from the white-light fit, with an uncertainty of 10\%. Finally, for the jitter, we assumed uniform values between 0 m/s and the best value coming from the white-light fit multiplied by a (arbitrary) factor of 10. When fitting the two transits simultaneously, $V_{sys}$ and $\sigma_W$ are still fit independently. Finally, the limb-darkening parameters were fixed independently for each wavelength range using the {\sc ldtk} package.

The minimum wavelength interval for which we can compute RV with the ESPRESSO DRS is the size of one spectral order. When fitting the two transits simultaneously, given that the precision on the RV (derived from the CCF) critically depends on the total S/N of the local spectrum
we decided to model the RM signal using two different wavelength binnings: one in steps of four spectral slices and the other in steps of eight slices. The resulting wavelength steps can be seen in Tables\,\ref{tab:transmission8} and \ref{tab:transmission4}. Due to the lower S/N of individual data, a binning of eight slices was used when fitting the transits individually. As mentioned above, due to telluric contamination some spectral orders have no CCF and were discarded. In any case, this implies that we do not have a regular spacing in wavelength.

\begin{table}[t]
\caption{Priors used in the chromatic light fit using {\sc Arome}. }
\begin{center}
\begin{tabular}{ll}
\hline
Parameter      & Prior\\
\hline
$V_{sys} [Km/s]$ & $\mathcal{U}(-15.27,-14.27)$\\
$R_p /R $ & $\mathcal{G}(0.1193,0.0119)$\\
$\sigma_W$ [m/s] &  $\mathcal{U}(0, 25)$\\
\hline
\hline
\end{tabular}
\label{tab:priorsChrom}
\end{center}
\end{table}

\begin{figure}[t!]
\begin{center}
\begin{tabular}{cc}
\includegraphics[angle=0,width=0.96\linewidth]{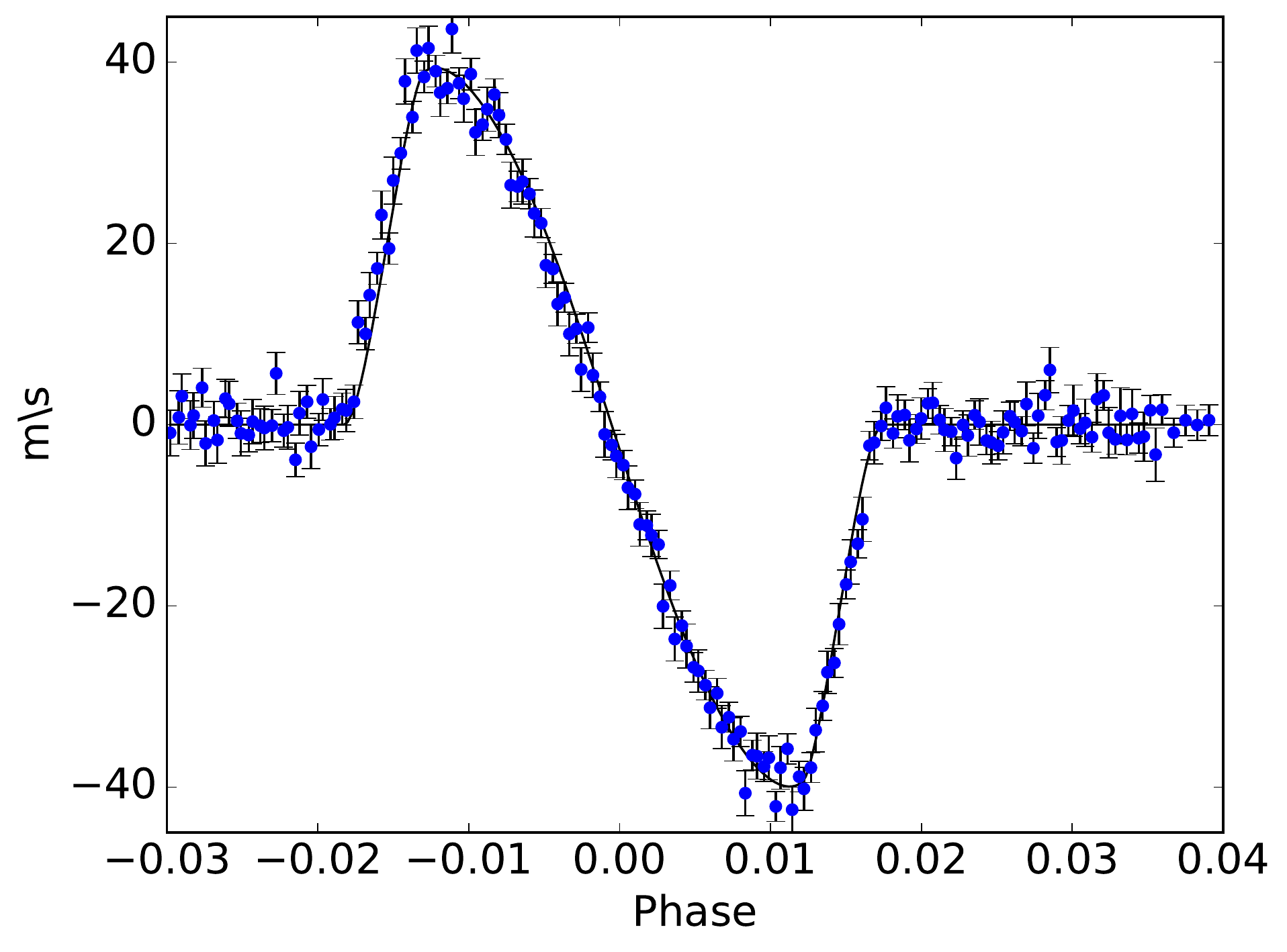}  \\
\includegraphics[angle=0,width=0.96\linewidth]{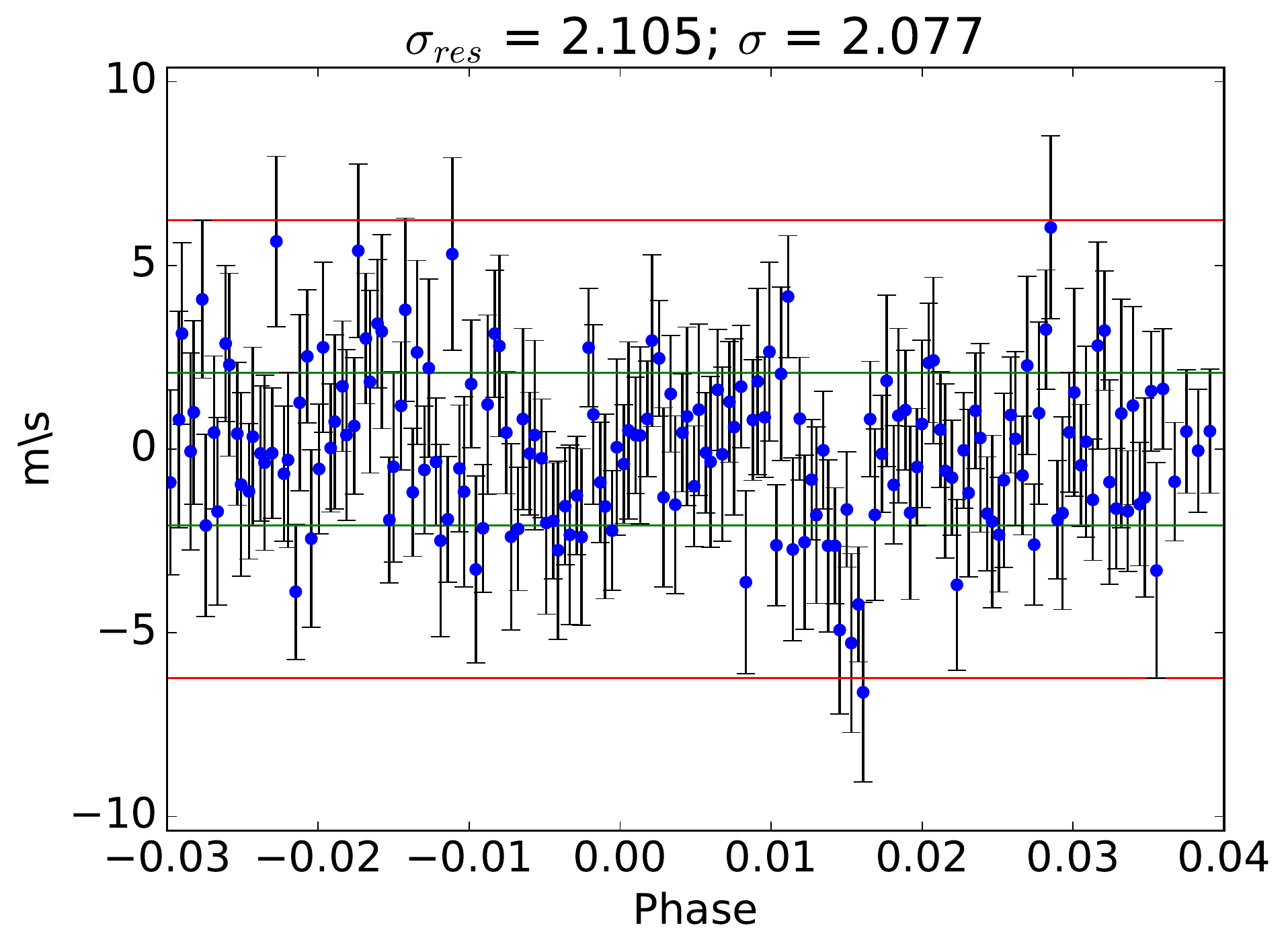}  \\
\end{tabular}
\end{center}
\vspace{-0.5cm}
\caption{Rossiter-McLaughlin fit to observed signal at 512\,nm (top) and residuals (bottom). Data from both transits are phase folded. The Keplerian function was subtracted from the data.}
\label{fig:RM512}
\end{figure}

\begin{figure*}[t!]
\begin{center}
\begin{tabular}{cc}
\includegraphics[angle=0,width=0.9\linewidth]{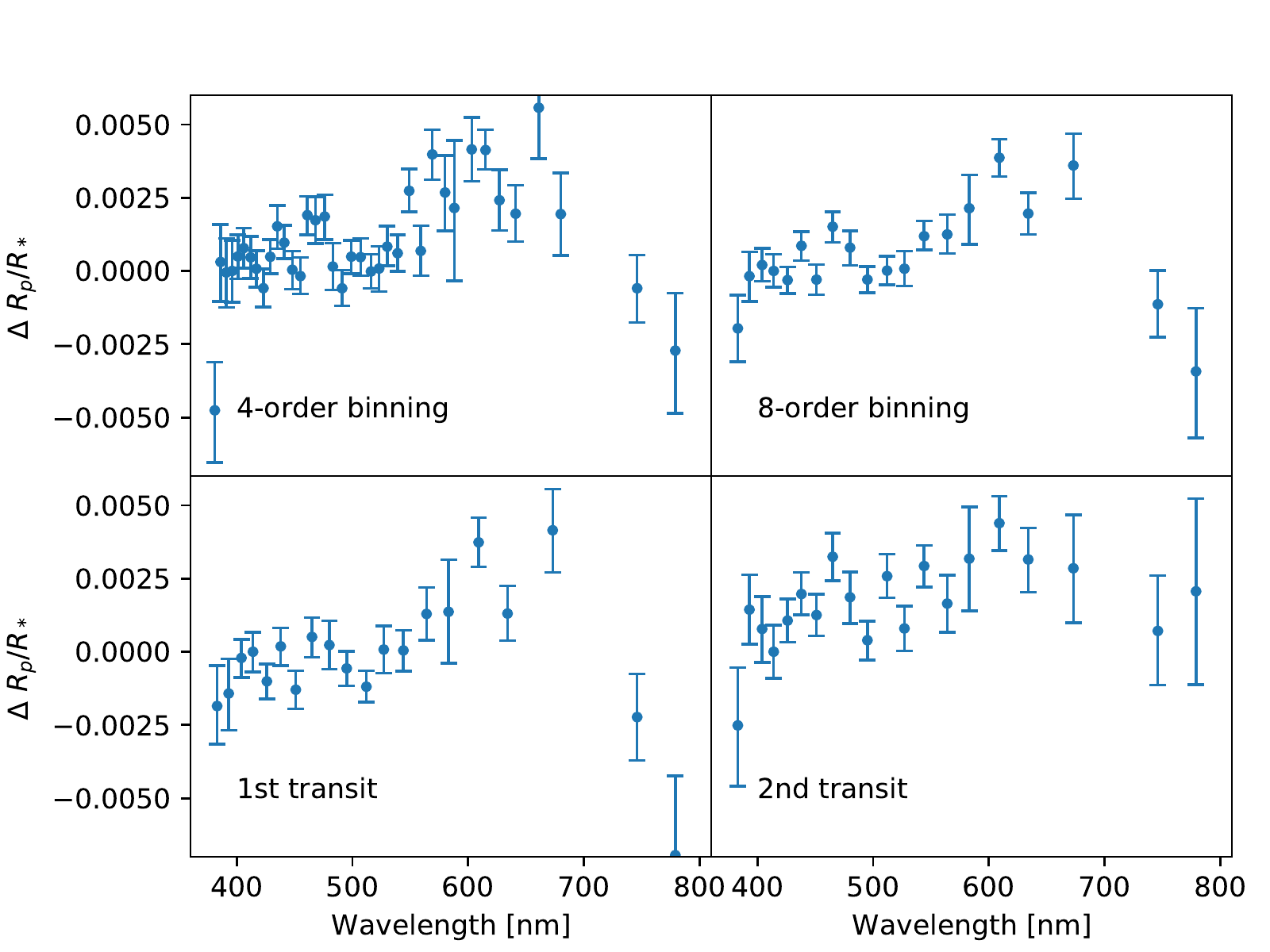} & 
\end{tabular}
\end{center}
\vspace{-0.5cm}
\caption{Transmission spectrum of HD209458 as obtained from the analysis of the chromatic RM curves. The two upper panels show the combined analysis of the two transits with four- and eight-slice binnings, respectively, for the left and right panels. The two lower panels present the analysis of the two transits separately as computed with an eight-slice binning.}
\label{fig:transmission}
\end{figure*}

For each wavelength range, the MCMC was run using a total of 30 chains taken randomly from the priors, allowing them to run for 2000 steps (burn-in) before continuing into the production chains (2000 steps). 
An example of an RM fit at 512\,nm, as well as the residuals, is presented in Fig.\,\ref{fig:RM512}.

The median value for the radius derived as a function of wavelength is then used to plot the transmission spectrum (Fig.\,\ref{fig:transmission}, see also Tables\,\ref{tab:transmission8} and \ref{tab:transmission4}). For clarity, we applied an (arbitrary) offset to the values of the radius ratio in all plots, subtracting the value measured in the fourth wavelength bin. In the top panels, we present the spectrum obtained by combining the two transits. The procedure was repeated by analysing each of the two transits separately (bottom panels). This allows us to check if the transmission spectrum we derived is similar in both cases. In all cases, the spectra show the same general trend: a relatively flat region below $\sim$500\,nm (0.5$\mu$m), and a bump starting around $\sim$550\,nm and finishing after $\sim$700\,nm. {Above 700\,nm, the two spectra show different transmission depths, with the second transit ``returning'' to values similar to the ones observed in the blue, while the first transit shows significantly lower transmission.} 

We note, however, that this trend, as it is obtained solely from the second transit, is less pronounced. 
It is not clear why we observe such a difference in the two transmission spectra as derived independently from the two transits separately. Part of the difference probably results from the lower (and more variable) S/N of the obtained spectra (see Fig.\,\ref{fig:obs}). As we see below, during the second night the star was also slightly more active (see next section). However, we cannot exclude instrumental effects or even real variability in the atmosphere of HD\,209458b, to explain the observed difference.

\subsection{Testing for robustness}

To confirm that the shape of the transmission spectrum did not depend on the RM model used, we performed two further tests. 

In the first one, we modeled the RM using the formalism of \citet[][]{Ohta-2005}, as implemented in {\sc PyAstronomy}. Though this code was built to use RVs derived from a template matching approach (opposite to a CCF method), its can still be used for a consistency check. As discussed above, this software only allows to use a linear limb darkening law. Comparing the two approaches also allows us to verify if the signal we are observing can be due to any wavelength dependent error in the limb darkening laws used.

\begin{figure}[t!]
\begin{center}
\begin{tabular}{c}
\includegraphics[angle=0,width=0.95\linewidth]{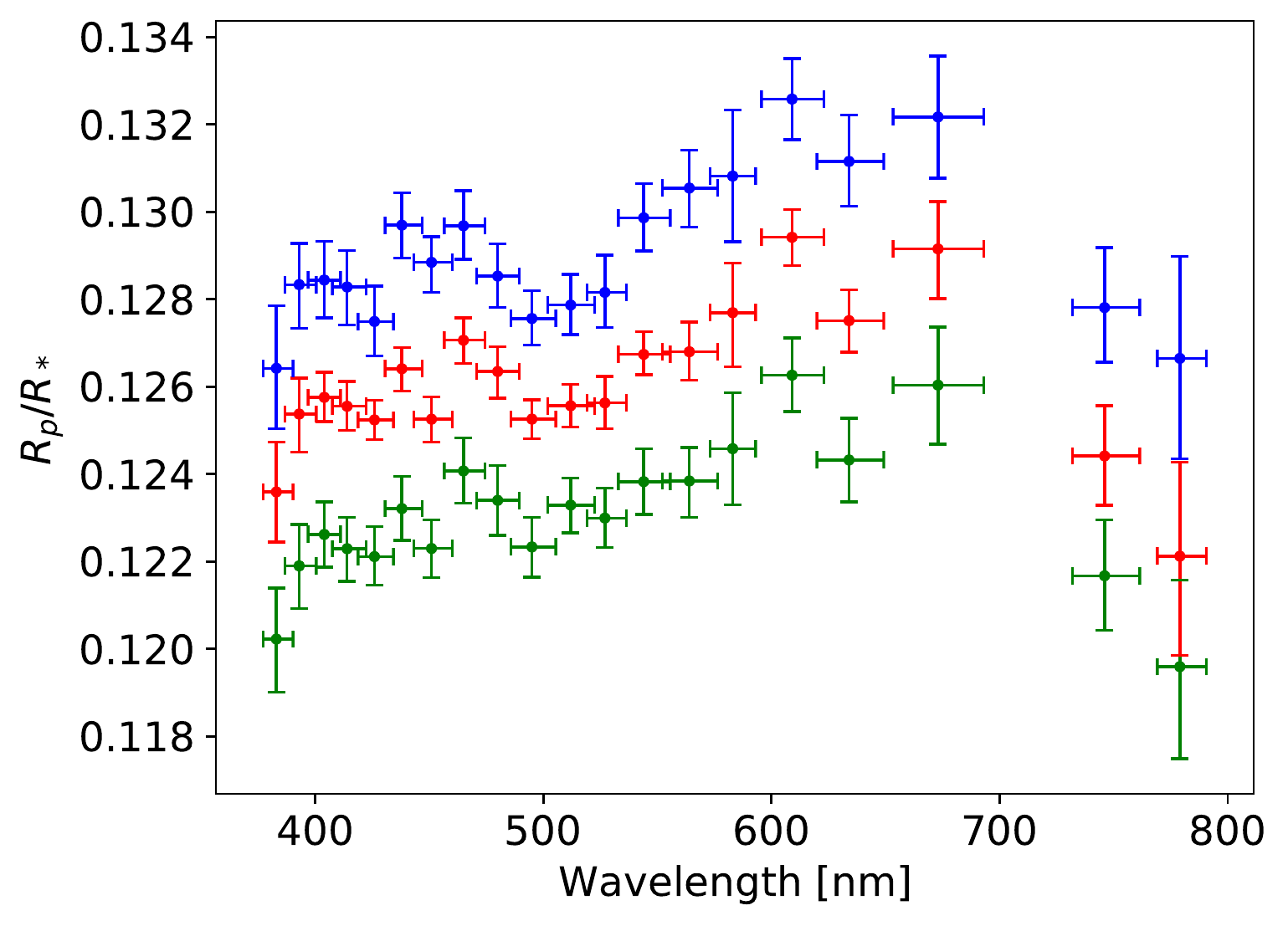}
\end{tabular}
\end{center}
\vspace{-0.5cm}
\caption{Comparison of derived transmission spectrum using the standard {\sc Arome} simulation (red), {\sc PyAstronomy} (green), and {\sc Arome} with free limb darkening (blue). For clarity, an arbitrary constant vertical shift was added to the different transmission spectra.}
\label{fig:comp_arome_ohta}
\end{figure}

All priors and parameters used are the same as used above with {\sc Arome}. Exceptions come from the fact that in this case we do not give as input $\beta_0$, $\sigma_0$, and $\zeta_t$, but we need to include $\Omega$, the angular rotational velocity of the star. We also assumed that the inclination between stellar spin axis and the y-axis is zero, meaning that the input $V_{rot}$ is equal to $v\sin{i}$.
In Fig.\,\ref{fig:comp_arome_ohta}, we present a comparison of the transmission spectrum obtained using the two formalisms (using the eight-slice binning). As is clear from the plot, the differences are minor, and always within the errors. We can thus conclude that the shape of the observed transmission spectrum does not depend on the method used to model the RM effect.\footnote{This is so even if the {\sc PyAstronomy} package is not optimized for the RM modeling of RVs derived using the cross-correlation method \citep[][]{Boue-2013}.} 

We further decided to redo the fit using the {\sc Arome} formalisms, but instead of fixing the (quadratic) limb darkening parameters, we let them vary freely for each wavelength bin using uniform priors between $0$ and $1$, for both the linear and quadratic terms.\footnote{We also tested with broader values between $-2$ and $+2$, but even with these nonphysical priors the results did not change.}
The values that were estimated from the white-light fit (0.62$\pm$0.13 and 0.36$\pm$0.22, respectively for the linear and quadratic term) are reasonably compatible, within the errors, with the ones derived with {\sc ldtk} (0.56 and 0.12, respectively). The limb-darkening coefficients did not converge in the chromatic curves (due to insufficient S/N of the data). More importantly, the obtained transmission spectrum is very similar to the one obtained with limb-darkening values fixed using {\sc ldtk} (Fig.\,\ref{fig:comp_arome_ohta}). 

\section{Discussion}
\label{sec:discussion}

\begin{figure*}[t!]
\begin{center}
\includegraphics[angle=0,width=0.9\linewidth]{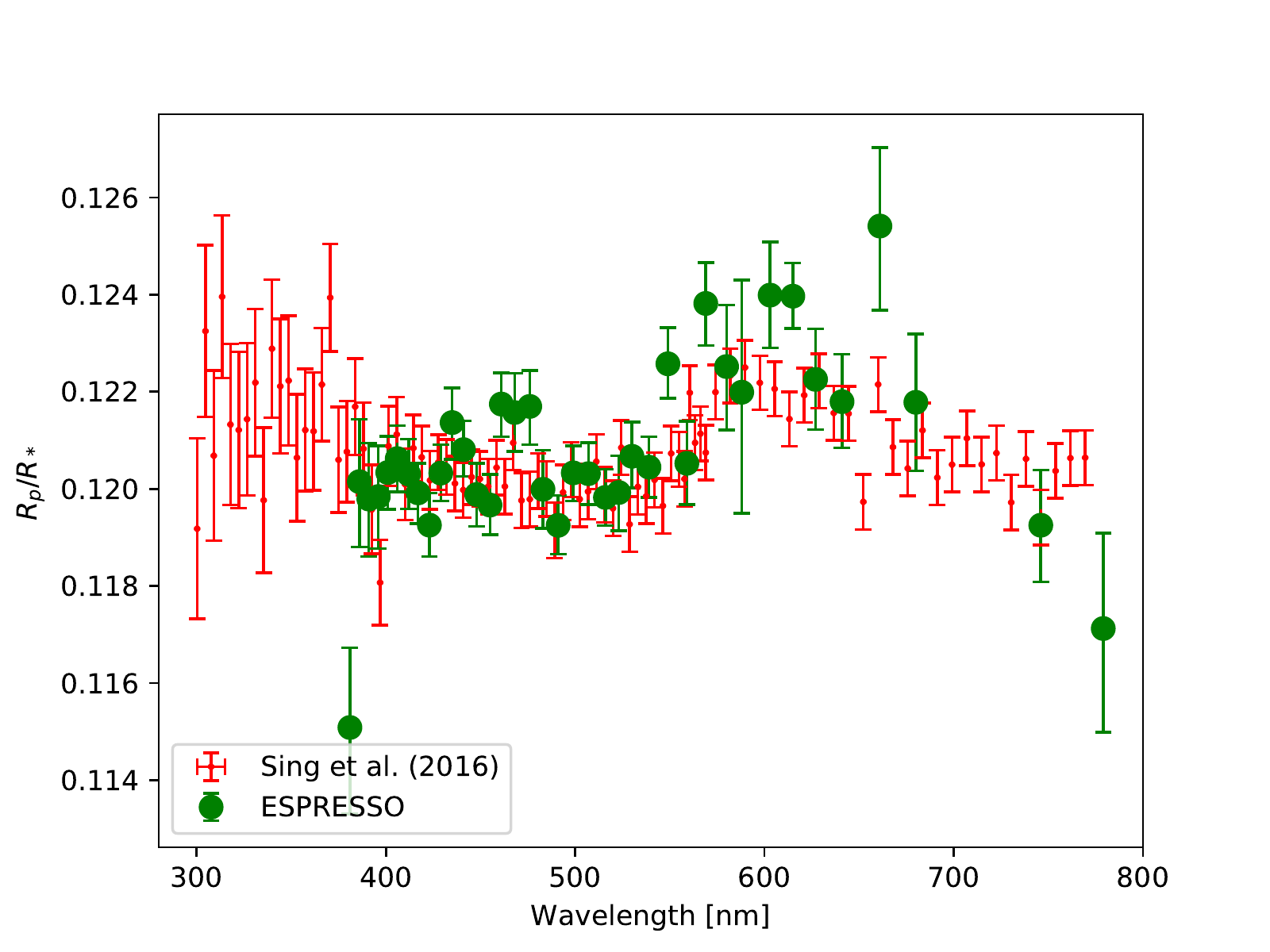}  \\
\end{center}
\vspace{-0.5cm}
\caption{Comparison of the observed transmission spectrum with the one obtained from HST observations and presented in \citet{Sing-2016}. {Green dots correspond to the four-order binning as presented in Fig.\,\ref{fig:transmission}.}}
\label{fig:compsing}
\end{figure*}

In Fig.\,\ref{fig:compsing}, we compare the transmission spectrum (four-slice binning) we obtained with one derived using HST data and presented in \citet[][]{Sing-2016}. The data from Sing et al. were compiled from the Exoplanet Atmospheric Spectral Library.\footnote{https://pages.jh.edu/$\sim$dsing3/David\_Sing/Spectral\_Library.html} We adopt the dataset version dubbed ``2016 Update preferred'.'\footnote{This version is issued from a more robust data analysis process using an up-to-date method in fitting the light curves and dealing with the systematics (D. Sing, Priv. Comm).} Due to our uncertainties in the derivation of ``absolute'' planet radii, in the plot we added an offset to the values of the radius ratio derived using our analysis (subtracting the average difference between our values and those in Sing's dataset).

A comparison of the two spectra indicates that both show similar trends, even if the ESPRESSO data has lower resolution (due to the spectral binning). In particular, we observe a clear increase in the radius above $\sim$550\,nm, an absorption that was identified as being due to Sodium \citep[][]{Sing-2008,Desert-2008}. Furthermore, the additional absorption redwards of $\sim$600\,nm, already discussed by these authors, is also clearly observed. \citet[][]{Desert-2008} suggested that this extra-absorption is possibly due to vanadium oxide (VO) and titanium oxide (TiO) in the atmosphere of HD\,209458b. Interestingly, the absorption peaks due to TiO, expected near 620 and 670\,nm (see Fig.\,3 in D\'esert et al.), roughly correspond to two of the peaks (higher radius) we observe in the ESPRESSO absorption spectrum. The latter of these two features does not appear in the HST data, and as we will see in Fig.\,\ref{fig:platon}, neither is it exactly at the expected wavelength. 

It is worth noting that with our dataset, the transmission spectroscopy study of Casasayas-Barris et al. (in prep.) did not manage to detect absorption lines of TiO or VO in high resolution. A similar result concerning TiO was also reported by \citet[][]{Hoeijmakers2015}. As highlighted in Hoeijmakers et al., however, the methods applied imply the use of cross-correlation ("stacking" all individual TiO or VO lines together), and are strongly dependent on the assumption about the expected (model) TiO and VO line spectra.

It is nevertheless interesting that, with an independent dataset and using a different methodology, we observe the same general trend (even if at a lower resolution) that was observed in the 2000 HST dataset. {The plot in Fig.\,\ref{fig:compsing} also suggests that the additional absorption redwards of $\sim$600\,nm may be slightly stronger in the ESPRESSO data when compared with the one observed in the HST dataset}. It would be interesting to obtain further observations to confirm if this is a signature of atmospheric variability \citep[see e.g.,][]{Parmentier2013}.

Due to telluric contamination, we were not able to obtain transmission data around 700\,nm. However, the two reddest wavelength bins show, again, lower values of the planet radius, compatible with the ones found near 500\,nm. {The reddest bin is, however, a bit lower. This point has large error bars and its value should be considered with care. First, its low level mostly comes from the data obtained during the first transit. Also, in the redder part of the spectrum, the number of stellar lines available to compute the RV decreases (the CCF mask used to compute the RV only uses 11 and 15 spectral lines for the two last orders, respectively, when compared to the $\sim$150 in the region near 500\,nm). This fact can by itself explain the higher error bars.}

It is also worth noting that the peak we {tentatively see} near 673\,nm is close to the Lithium line region \citep[670,7\,nm -- see Fig.\,1 and model of][]{Desert-2008}. We note, however, that such feature was not observed in the HST data. Near 460\,nm, we also see a small increase in the derived radius, in a region where HST data also shows a small increase (in one single wavelength bin).

The bluest bin in our transmission spectrum is significantly below the remaining ones, and incompatible with the HST data. As we can see from Fig.\,\ref{fig:transmission}, this decrease is only seen in the second transit. Such behavior could be explained by a spot crossing event or if HD\,209458 had a faculae-richer surface by the time of the second transit \citep[][]{Oshagh-2014,Boldt2020}. 
An estimate of the activity level of the star using the \ion{Ca}{ii} H and K lines indeed shows a small (but significant) increase in the activity level of the star from the first to the second transit (the value of $\log{R'_{HK}}$ changed from $-4.850\pm0.003$ to $-4.840\pm$0.002).\footnote{Values represent the median and standard deviation of all the measurements obtained from individual spectra in each night.} We should note, however, that this variation is only marginal. The lower S/N in the observed bluest part of the spectrum, with usual values below 30 for the second night (and descending to below ten for the last measurements) is likely a better explanation for the observed discrepancy. Systematic (though difficult to quantify) effects may indeed be expected as we approach the read-out noise limit.

\subsection{Modeling with PLATON}

We decided to model the observed transmission spectrum of HD\,209458b using the publicly available PLATON code \citep[][]{Zhang-2019,Zhang-2020}, assuming the recommended prescription with an isothermal atmosphere and equilibrium chemistry. We used the MultiNest approach (with 1000 live points) to fit the data, and used the example that is provided for HD\,209458b as a baseline together with the code \citep[see][]{Zhang-2019}. We only changed the estimated values of the stellar radius (R$_\star=1.162R{_\odot}$) and the planet mass ($M_p=0.714\,M_{Jup}$), both from TEP-Cat \citep[][]{Southworth2010}, and the effective temperature of the star, using the value already applied in the RM analysis. A solar C/O ratio of 0.53 was also used; we expect this to be a good approximation given that HD\,209458 has solar metallicity \citep[][]{Sousa2008,Suarez-Andres-2018}. 
In the fitting procedure, the priors were defined as follows. For $R_\star$ and $M_p,$ we used Gaussian priors centered in the above-mentioned values and with widths of 0.02$R_\odot$ and 0.04$M_{jup}$, respectively. For the remaining parameters, we used uniform priors: 
$0.9\times1.38<R_p<1.1\times1.38\,R_{jup}$, 
$0.5\times1\,459<T_p<1.5\times1\,459\,K$\footnote{The central value for the equilibrium temperature used was taken from TEP-Cat.},
$-0.99<\log{P_{top}}<8$,
$-1<\log{Z}<3$, a multiplicative error factor between 0.5 and 5,
and a logarithm of the scattering factor between 0 and 2.
We also defined the scattering slope using a pure Rayleigh scattering model, as recommended.

We then tried to fit the data including (or excluding) Na, TiO, and VO, following the example of \citet[][]{Desert-2008}. A model with no absorption species was also tested as a baseline. Given the uncertainties in the bluest and reddest bins in our data (see above), we decided to exclude these two points from the analysis. 

\begin{figure*}[t!]
\begin{center}
\begin{tabular}{cc}
\includegraphics[angle=0,width=0.47\linewidth]{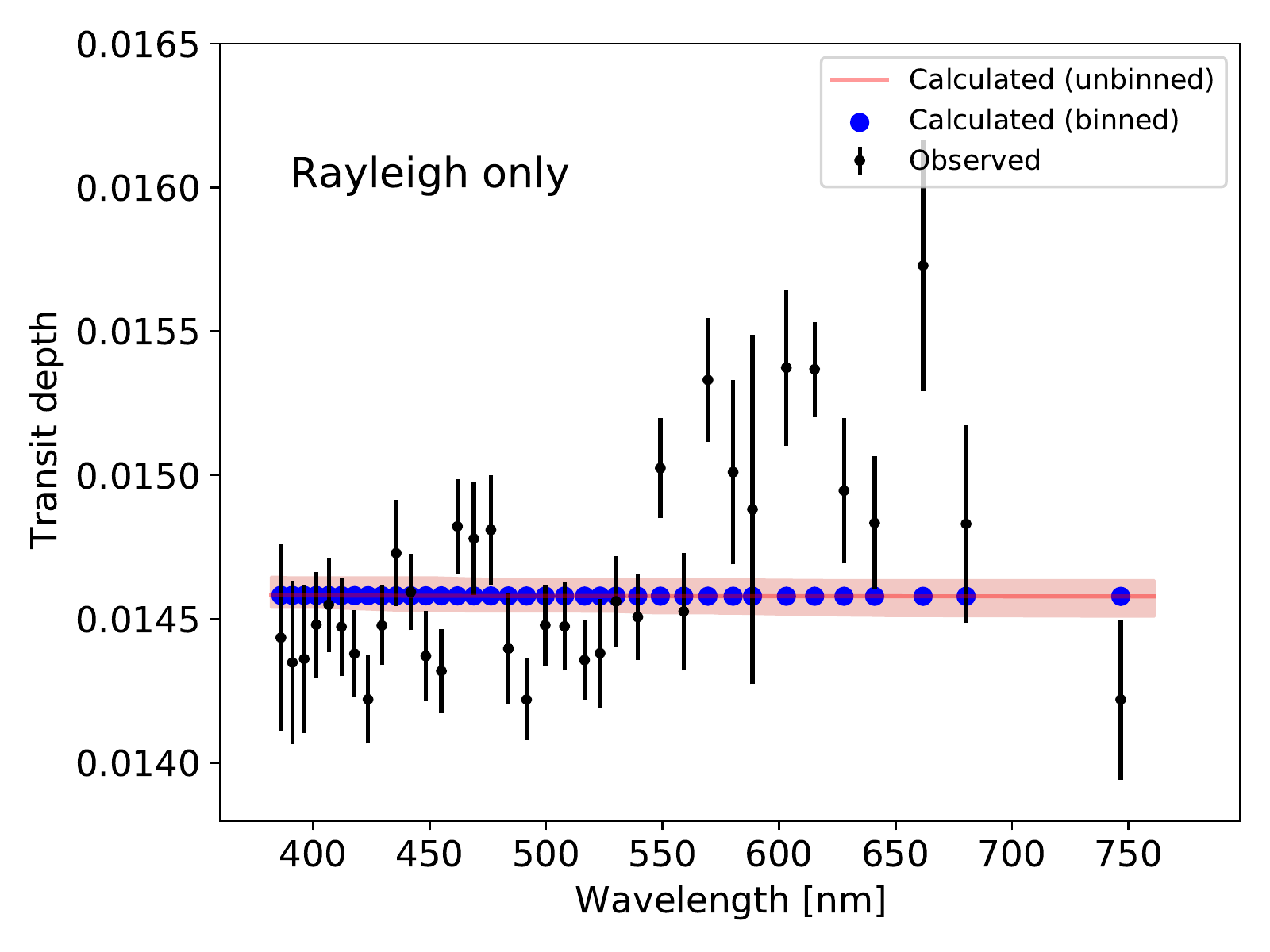} &
\includegraphics[angle=0,width=0.47\linewidth]{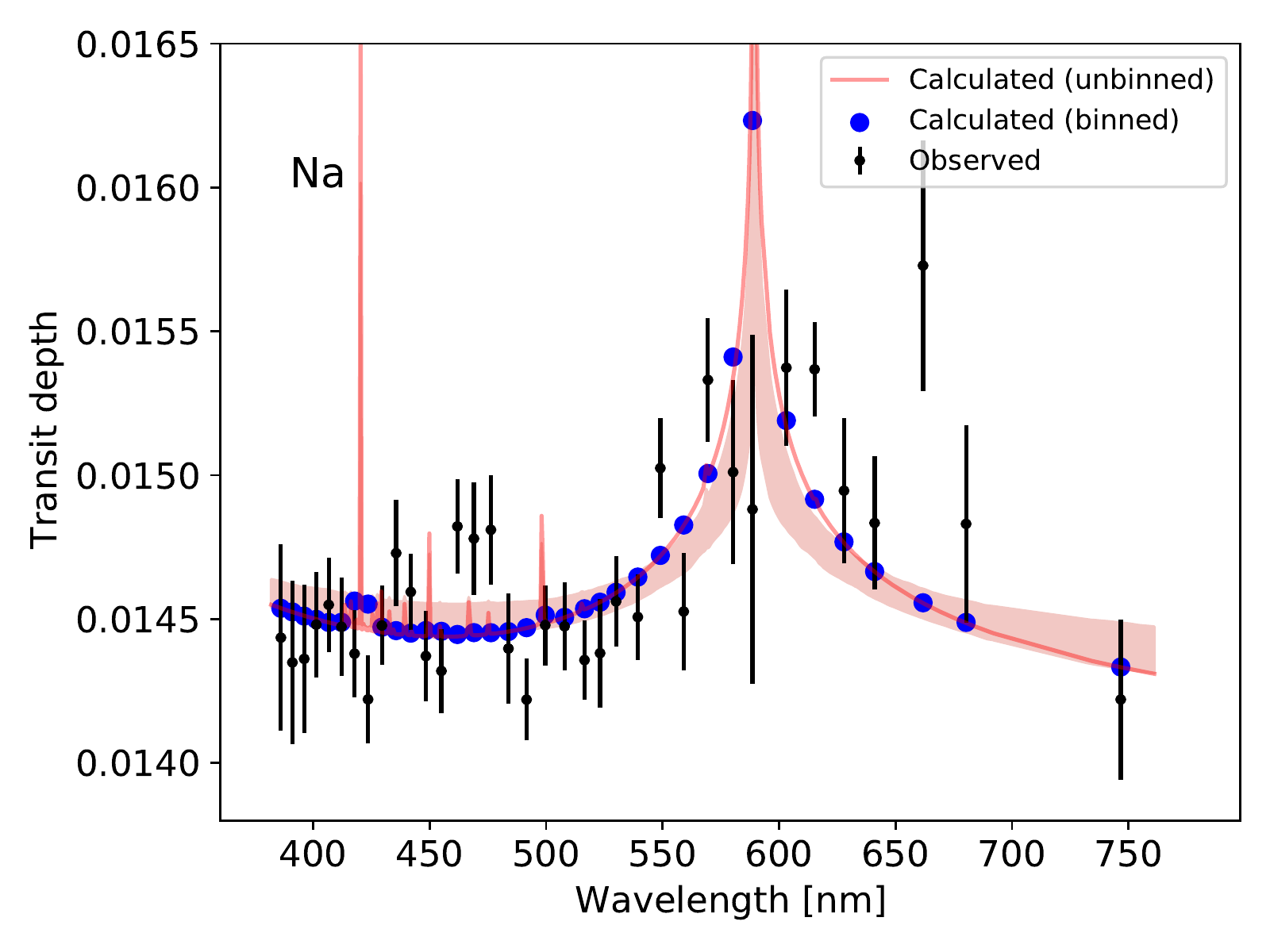}\\
\includegraphics[angle=0,width=0.47\linewidth]{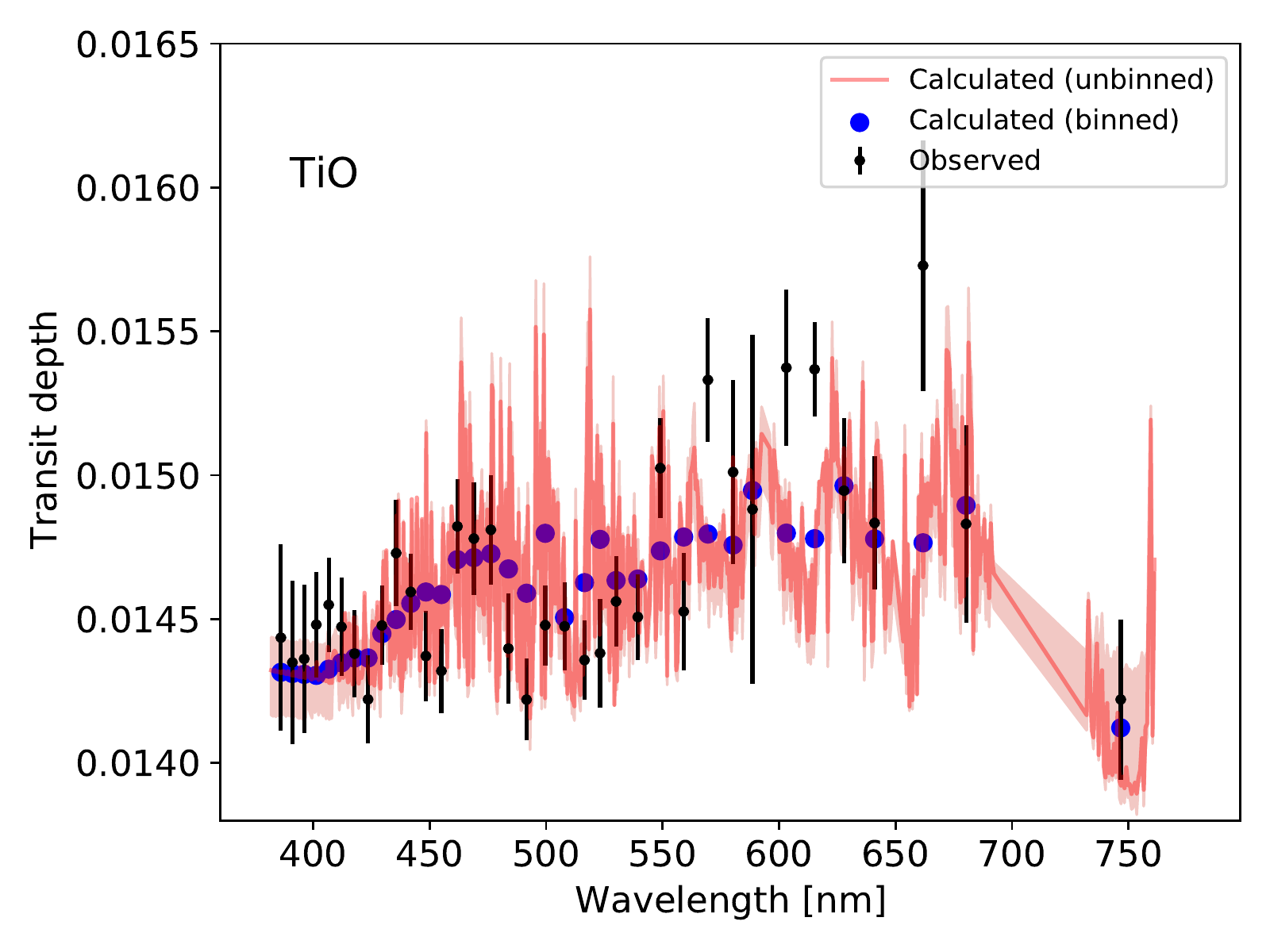} &
\includegraphics[angle=0,width=0.47\linewidth]{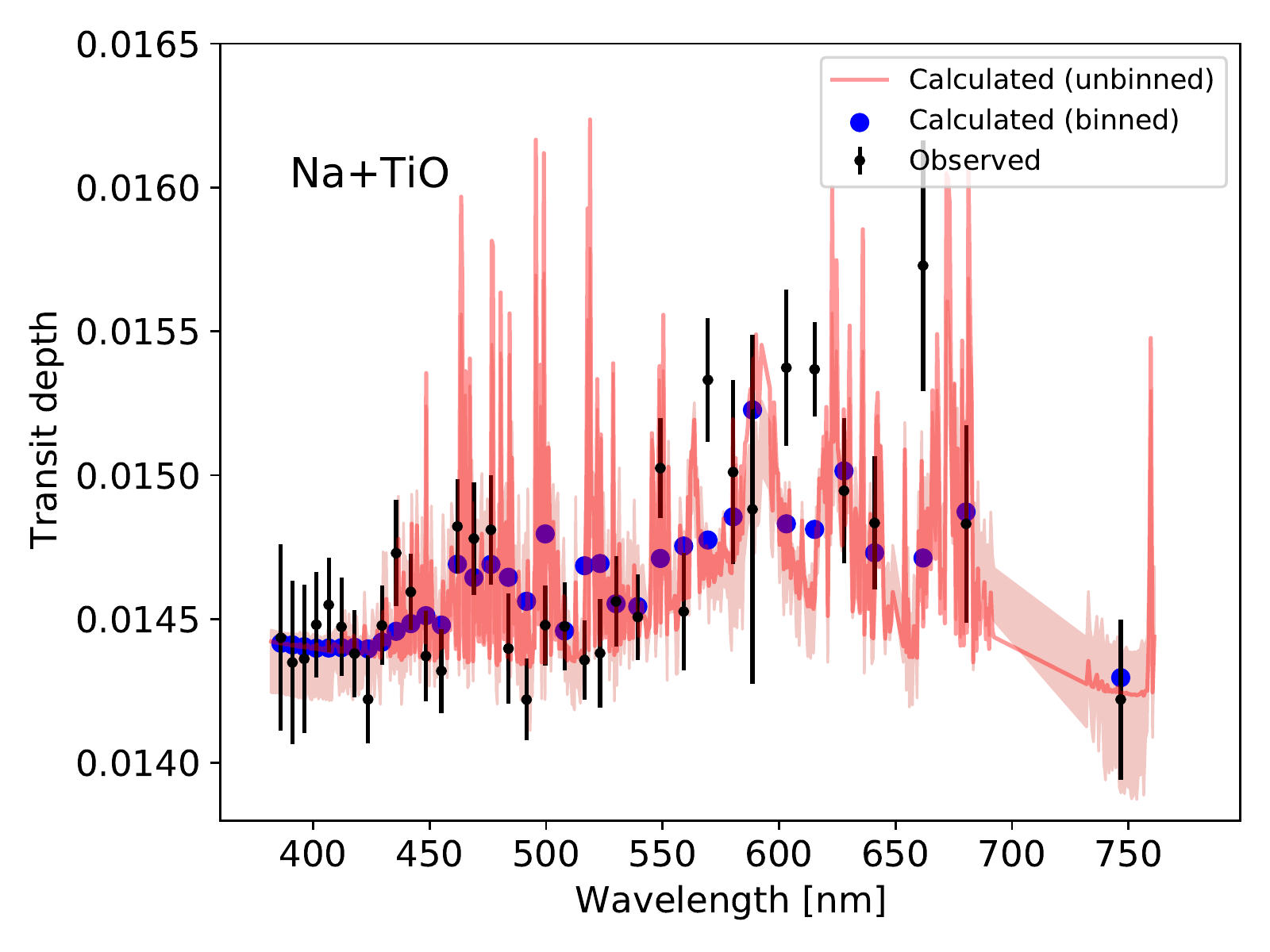}\\
\includegraphics[angle=0,width=0.47\linewidth]{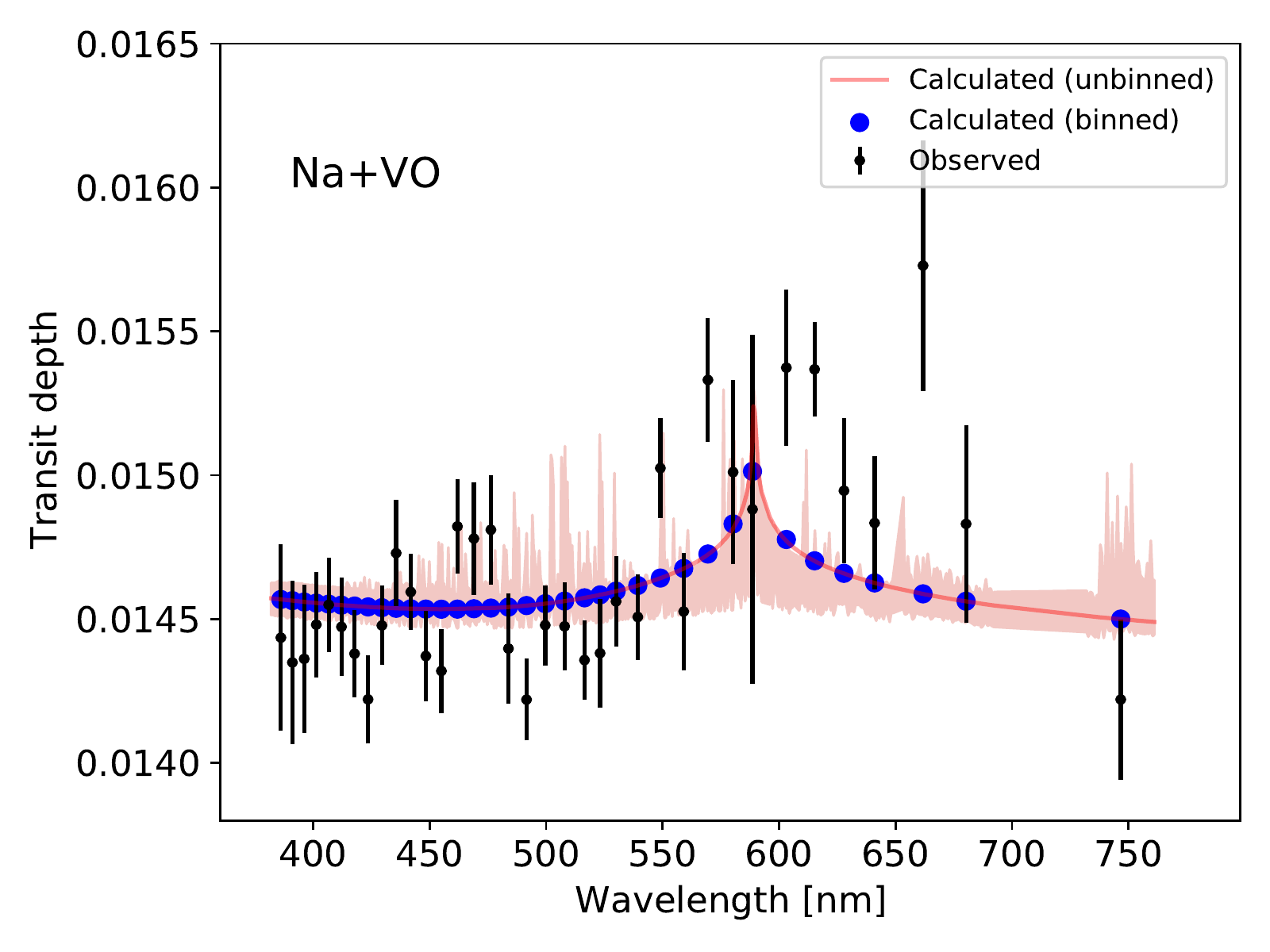} &
\includegraphics[angle=0,width=0.47\linewidth]{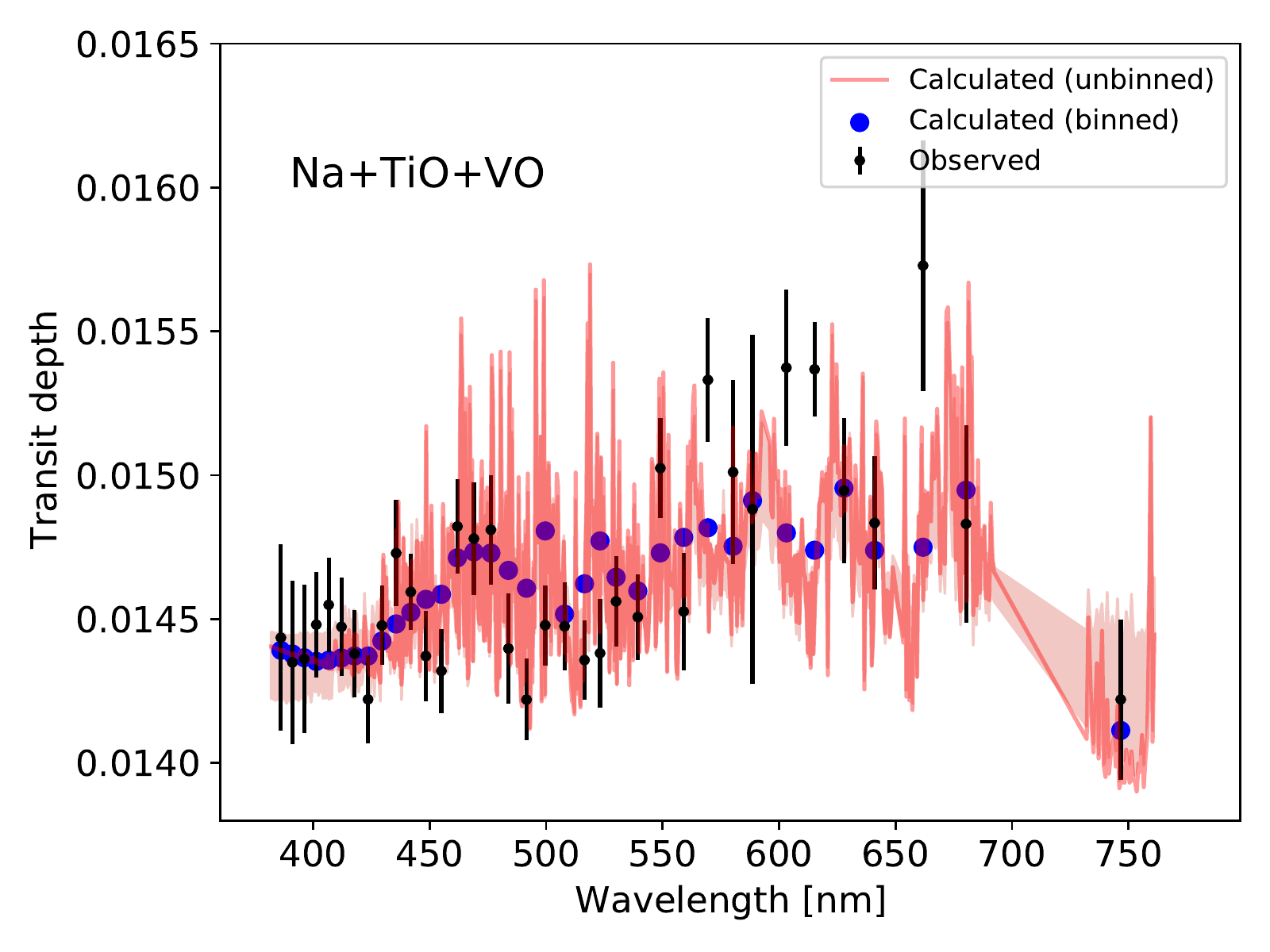}\\
\end{tabular}
\end{center}
\vspace{-0.5cm}
\caption{Comparison of our transmission spectrum with {synthetic models} computed using PLATON. Top: model with ``no elements'' (left) and with only Na absorption (right). Middle: models with only TiO (left) and with Na + TiO (right). Bottom: models with Na + VO (left) and with all Na, VO, and TiO (right). In all plots, the observational points are denoted in black, with the respective error bars. The red line denotes the best high-spectral resolution model, while the blue points represent its wavelength binned version at the same resolution as the observations. The red shaded area represents the sixteenth and eighty-fourth percentiles of the models.}
\label{fig:platon}
\end{figure*}

Figure\,\ref{fig:platon} presents some of our results. In Table\,\ref{tab:bayes}, we also present the natural logarithm of the Bayesian evidence ($\ln{z}$) of the different models tested. We assume that a model is preferred if this value increases by at least three \citep[][]{Kass1995}. As can be seen from the Table, models assuming only the Na, TiO, or a combination of the two, as well as the model with Na, TiO, and VO, all have statistically undistinguishable $\ln{z}$ values. However, all these models are clearly preferred over the ``no-elements'' model, or the models with only VO or Na+VO. 

\begin{table}[b]
\caption{Logarithm of the Bayesian evidence ($\ln{z}$) of the different models fit with PLATON. Models are ordered by increasing evidence.}
\begin{center}
\begin{tabular}{ll}
\hline
Model  & $\ln{z}$\\
\hline
No elements & 230.2$\pm$0.1\\
Only VO & 230.3$\pm$0.1\\
Na + VO & 231.5$\pm$0.2 \\
Na + TiO + VO & 234.1$\pm$0.1\\
Only Na & 234.3$\pm$0.2\\
Only TiO & 234.4$\pm$0.1\\
Na + TiO & 234.6$\pm$0.1\\ 
\hline
\hline
\end{tabular}
\label{tab:bayes}
\end{center}
\end{table}

It is also interesting to see that the best model (highest $\ln{z}$) is the one that assumes Na and TiO. The bump observed near 480\,nm is reasonably well reproduced by a model including TiO absorption. Interestingly, TiO alone is also able to explain, to the precision level of our data (and to the limits of the model used) the observed transmission spectrum. On the other hand, we find no evidence for the presence of VO absorption.

We should note, however, that none of the models is fully capable of reproducing our dataset, in the sense that in the regions above $\sim$600\,nm, our data show an absorption level above the modeled one. The observed mismatch may be caused by modeling approximations (e.g., the isothermal T-P profile), observational errors, effects of stellar activity \citep[e.g.,][]{Sedaghati2017,Espinoza2019}, or by the presence of additional absorbing species such as CaH or ZrO \citep[][]{VanEck2017}. 

In a recent paper, \citet[][]{Casasayas-Barris-2020} could not confirm the previous detections of Na absorption as measured at the Na-doublet line cores. They attribute the previous claims to the unaccounted-for Rossiter-McLaughlin effects in the modeling of line profiles. This result has also been confirmed using our ESPRESSO data (Casasayas-Barris et al., in prep.). It may at first sound intriguing that the Na wings may have been detected, while no detectable (within the errors of the ESPRESSO observations and analysis) absorption is measured at the line cores.\footnote{Absorption can however be masked by the dominant RM effect changing the shape of the stellar lines.} Although not conclusive, our result could reinforce the idea that TiO alone may be observed, and that no Na is detected in the atmosphere of HD\,209458. 

{We should note, however, that the methods used in \citet[][]{Casasayas-Barris-2020} and in this paper probe the transmission spectrum at very different spectral resolutions. \citet[][]{Casasayas-Barris-2020} focus their analysis on the detection of the core of the Na line, and the methodology used does not make it possible to probe broadband variability. On the other hand, in the present study, we are probing broadband variability, thus being sensitive to variations on scales comparable to the expected wings of the Na spectral lines; we are not sensitive to the line core.}

Indeed, we could alternatively speculate that the sodium cores may be formed so high in the atmosphere that all Na is ionized, while the wings are formed at a lower temperature region. This could be evidence for the existence of a thermal inversion in the atmosphere of HD\,209458b. Although evidence for significant thermal inversion in HD\,209458b has not been detected using NIR spectroscopy \citep[e.g.,][]{Schwarz2015}, it is interesting to add that TiO has been suggested as a possible source of thermal inversions \citep[][]{Desert-2008,Parmentier2013,Piette2020}.

To try to test this latter scenario, we decided to rerun PLATON for the models with only Na, only TiO, and Na + TiO, after excluding the two data points near the core of the Na lines. The results show that the $\ln{z,}$ the three models are very similar and statistically undistinguishable. This test is thus inconclusive.

\section{Conclusions}
\label{sec:conclusions}

In this paper, we used ESPRESSO high-resolution spectra to extract the broadband transmission spectrum of HD\,209458b, using the chromatic Rossiter-McLaughlin method. We show that the obtained results are compatible with those derived in previous studies using HST data. 

We modeled the obtained transmission spectrum using the PLATON software \citep[][]{Zhang-2020}. The results suggest that, to the precision level of the data and modeling, either Na+TiO and Na or TiO alone can explain the observed transmission spectrum. {However, none of the generated models were fully capable of explaining the observed dataset.} More data and a more detailed model may be needed to clarify the exact chemical species responsible for the observed spectrum.

The results presented in this paper show that the application of the chromatic Rossiter-McLaughlin method to ESPRESSO data can be used to set  important constraints on the transmission spectrum of exoplanets. With just two transits, we were able to obtain a transmission spectrum of HD\,209458b that shows the same structures, albeit at lower resolution, as the one obtained with four HST transits. It will be interesting to apply the same methodology to other targets observed with ESPRESSO.

\begin{acknowledgements}

To my grandparents. The authors acknowledge the ESPRESSO project team for its effort and dedication in building the ESPRESSO instrument. This work was supported by FCT - Funda\c{c}\~ao para a Ci\^encia e a Tecnologia through national funds and by FEDER through COMPETE2020 - Programa Operacional Competitividade e Internacionaliza\c{c}\~ao by these grants: UID/FIS/04434/2019; UIDB/04434/2020; UIDP/04434/2020; PTDC/FIS-AST/32113/2017 \& POCI-01-0145-FEDER-032113; PTDC/FIS-AST/28953/2017 \& POCI-01-0145-FEDER-028953; PTDC/FIS-AST/28987/2017 \& POCI-01-0145-FEDER-028987. NJN and VA acknowledge support by FCT through projects reference PTDC/FIS-OUT/29048/2017, IF/00852/2015, and IF/00650/2015/CP1273/CT0001.
The INAF authors acknowledge financial support of the Italian Ministry of Education, University, and Research
with PRIN 201278X4FL and the "Progetti Premiali" funding scheme. J. H. C. M. is supported in the form of a work contract funded by national funds through FCT (DL 57/2016/CP1364/CT0007). This work has been carried out within the framework of the National Centre of Competence in Research PlanetS supported by the Swiss National Science Foundation. R.A. acknowledge the financial support of the SNSF.
This project has received funding from the European Research Council (ERC) under the European Union's Horizon 2020 research and innovation programme (project Four Aces grant agreement No 724427). SCCB acknowledges support from  Funda\c{c}\~ao para a Ci\^encia e a Tecnologia (FCT) through Investigador FCT contract IF/01312/2014/CP1215/CT0004. J.P.F. is supported in the form of a work contract funded by national funds through FCT with reference DL57/2016/CP1364/CT0005. ASM acknowledges financial support from the Spanish Ministry of Science and Innovation (MICINN) under the 2019 Juan de la Cierva Programme. A.S.M. acknowledge financial support from the Spanish Ministry of Science, Innovation and Universities (MICIU) AYA2017-86389-P.
XD is grateful to The Branco Weiss Fellowship--Society in Science for its financial support. 
This project has received funding from the European Research Council (ERC) under the European Union's Horizon 2020 research and innovation programme (project SCORE grant agreement No 851555)
This project has received funding from the European Research Council (ERC) under the European Union's Horizon 2020 research and innovation programme (project {\sc Four Aces}; grant agreement No 724427). S.G.S acknowledges the support from FCT through Investigador FCT contract nr. CEECIND/00826/2018 and POPH/FSE (EC).

\end{acknowledgements}

\bibliographystyle{aa}
\bibliography{santos_bibliography}

\appendix

\section{Corner plot for the white-light curve}

\begin{figure*}[t!]
\begin{center}
\includegraphics[angle=0,width=0.9\linewidth]{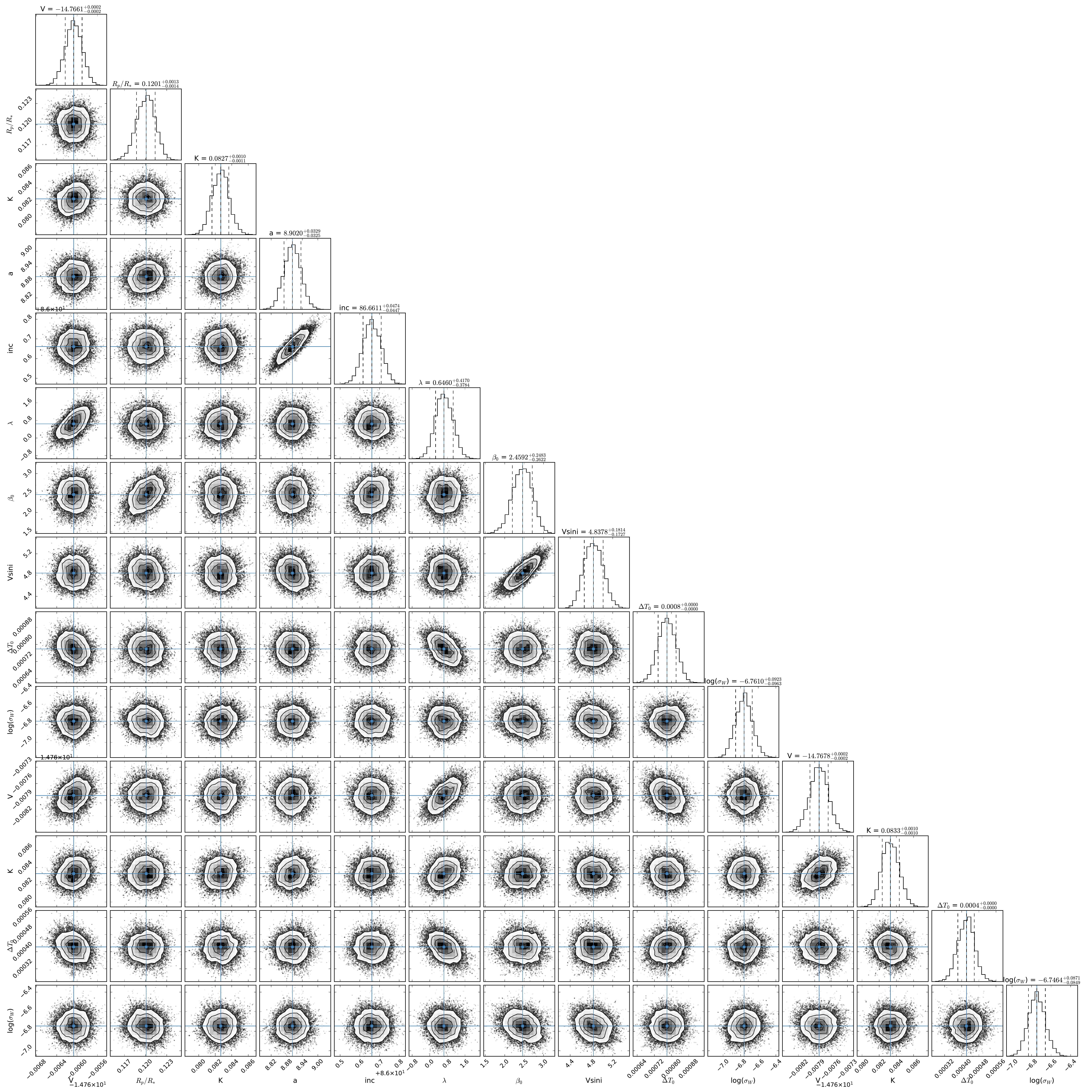}
\end{center}
\vspace{-0.5cm}
\caption{Corner plot of white-light fit to the two transits. The values and error bars represent the median and 1-sigma uncertainty based on the posterior distributions. The lower right plots for $V_{sys}$, $\Delta \phi_0$, $K$, and $\sigma_W$ concern the second transit.}
\label{fig:cornerWL}
\end{figure*}

\section{Data with transmission curves}

\begin{table}[t]
\caption{Planetary radius (in stellar radius units) as a function of wavelength (transmission spectrum) as obtained from our analysis (eight-slice binning). }
\begin{center}
\begin{tabular}{ll}
\hline
Central $\lambda$  & $R_p/R_\star$\\
(nm) & \\
\hline
383 & 0.1206$^{+0.0011}_{-0.0011}$\\
393 & 0.1224$^{+0.0008}_{-0.0009}$\\
404 & 0.1228$^{+0.0006}_{-0.0006}$\\
414 & 0.1226$^{+0.0006}_{-0.0006}$\\
426 & 0.1222$^{+0.0005}_{-0.0005}$\\
438 & 0.1234$^{+0.0005}_{-0.0005}$\\
451 & 0.1223$^{+0.0005}_{-0.0005}$\\
465 & 0.1241$^{+0.0005}_{-0.0005}$\\
480 & 0.1233$^{+0.0006}_{-0.0006}$\\
495 & 0.1223$^{+0.0004}_{-0.0004}$\\
512 & 0.1226$^{+0.0005}_{-0.0005}$\\
527 & 0.1226$^{+0.0006}_{-0.0006}$\\
544 & 0.1237$^{+0.0005}_{-0.0005}$\\
564 & 0.1238$^{+0.0007}_{-0.0006}$\\
583 & 0.1247$^{+0.0012}_{-0.0012}$\\
609 & 0.1264$^{+0.0006}_{-0.0006}$\\
634 & 0.1245$^{+0.0007}_{-0.0007}$\\
673 & 0.1261$^{+0.0011}_{-0.0011}$\\
746 & 0.1214$^{+0.0012}_{-0.0011}$\\
779 & 0.1191$^{+0.0021}_{-0.0023}$\\
\hline
\hline
\end{tabular}
\label{tab:transmission8}
\end{center}
\end{table}

\begin{table}[t]
\caption{Planetary radius (in stellar radius units) as a function of wavelength (transmission spectrum), as obtained from our analysis (four-slice binning). }
\begin{center}
\begin{tabular}{ll}
\hline
Central $\lambda$  & $R_p/R_\star$\\
(nm) & \\
\hline
381 & 0.1151$^{+0.0018}_{-0.0016}$\\
386  & 0.1201$^{+0.0013}_{-0.0013}$\\
391  & 0.1198$^{+0.0012}_{-0.0012}$\\
396  & 0.1198$^{+0.0011}_{-0.0010}$\\
401 &  0.1203$^{+0.0008}_{-0.0007}$\\
406  & 0.1206$^{+0.0007}_{-0.0007}$\\
412  & 0.1203$^{+0.0007}_{-0.0007}$\\
417  & 0.1199$^{+0.0006}_{-0.0006}$\\
423  & 0.1193$^{+0.0006}_{-0.0007}$\\
429  & 0.1203$^{+0.0006}_{-0.0006}$\\
435  & 0.1214$^{+0.0008}_{-0.0007}$\\
441  & 0.1208$^{+0.0005}_{-0.0006}$\\
448  & 0.1199$^{+0.0007}_{-0.0006}$\\
455  & 0.1197$^{+0.0006}_{-0.0006}$\\
461  & 0.1217$^{+0.0007}_{-0.0006}$\\
468  & 0.1216$^{+0.0008}_{-0.0008}$\\
476  & 0.1217$^{+0.0008}_{-0.0007}$\\
483  & 0.1200$^{+0.0008}_{-0.0008}$\\
491 &  0.1192$^{+0.0006}_{-0.0006}$\\
499  & 0.1203$^{+0.0006}_{-0.0006}$\\
507  & 0.1203$^{+0.0006}_{-0.0006}$\\
516  & 0.1198$^{+0.0006}_{-0.0006}$\\
523  & 0.1199$^{+0.0008}_{-0.0008}$\\
530  & 0.1207$^{+0.0006}_{-0.0007}$\\
539  & 0.1204$^{+0.0006}_{-0.0006}$\\
549  & 0.1226$^{+0.0007}_{-0.0007}$\\
559  & 0.1205$^{+0.0008}_{-0.0009}$\\
569  & 0.1238$^{+0.0009}_{-0.0008}$\\
580  & 0.1225$^{+0.0013}_{-0.0013}$\\
588  & 0.1220$^{+0.0025}_{-0.0023}$\\
603  & 0.1240$^{+0.0011}_{-0.0011}$\\
615  & 0.1240$^{+0.0007}_{-0.0007}$\\
627  & 0.1223$^{+0.0010}_{-0.0010}$\\
641  & 0.1218$^{+0.0009}_{-0.0010}$\\
661 &  0.1254$^{+0.0017}_{-0.0016}$\\
680  & 0.1218$^{+0.0014}_{-0.0014}$\\
746  & 0.1193$^{+0.0012}_{-0.0011}$\\
779  & 0.1171$^{+0.0021}_{-0.0020}$\\
\hline
\hline
\end{tabular}
\label{tab:transmission4}
\end{center}
\end{table}

\end{document}